\documentclass[aps,amsfonts,amssymb,nofootinbib,superscriptaddress]{revtex4}
\usepackage{graphics}
\usepackage{hyperref}
\usepackage{epsfig}
\usepackage{color}
\usepackage{amsmath}
\usepackage{braket}

\usepackage[version=3]{mhchem}
\usepackage{siunitx}
\usepackage{multirow}
\usepackage{bm}

\def\ket#1{|#1\rangle }
\def\bra#1{\langle #1 |}
\def\braket#1{\langle #1 \rangle}
\def\n{\nonumber \\ }

\usepackage[normalem]{ulem}  

\def\ket#1{|#1\rangle }
\def\bra#1{\langle #1 |}
\def\braket#1{\langle #1 \rangle}
\def\n{\nonumber \\ }

\makeatletter
\let\cat@comma@active\@empty
\makeatother

\usepackage[normalem]{ulem}  

\newcommand{\st}[1]{\left\{ #1 \right\}}

\renewcommand{\v}[1]{\boldsymbol{#1}}

\newcommand{\tr}{\operatorname{Tr}}
\newcommand{\im}{\operatorname{Im}}
\newcommand{\re}{\operatorname{Re}}

\DeclareMathOperator\erfc{erfc}

\newcommand{\ibz}{\int [d \v{k}]}

\everymath{\displaystyle}

\begin{document}

\title{Resonance-enhanced optical nonlinearity in the Weyl semimetal TaAs}



\author{Shreyas Patankar}
\affiliation{Department of Physics, University of California, Berkeley, California 94720, USA}
\affiliation{Materials Science Division, Lawrence Berkeley National Laboratory, Berkeley, California 94720, USA}
\author{Liang Wu}
\affiliation{Department of Physics, University of California, Berkeley, California 94720, USA}
\affiliation{Materials Science Division, Lawrence Berkeley National Laboratory, Berkeley, California 94720, USA}
\affiliation{Department of Physics \& Astronomy, University of Pennsylvania, Philadelphia, PA 19104}
\author{Baozhu Lu}
\affiliation{Department of Physics, Temple University, Philadelphia, Pennsylvania 19122, USA}
\affiliation{Temple Materials Institute, Temple University, Philadelphia, PA 19122, USA}
\author{Manita Rai}
\affiliation{Department of Physics, Temple University, Philadelphia, Pennsylvania 19122, USA}
\affiliation{Temple Materials Institute, Temple University, Philadelphia, PA 19122, USA}
\author{Jason D. Tran}
\affiliation{Department of Physics, Temple University, Philadelphia, Pennsylvania 19122, USA}
\affiliation{Temple Materials Institute, Temple University, Philadelphia, PA 19122, USA}
\author{T. Morimoto}
\affiliation{Department of Physics, University of California, Berkeley, California 94720, USA}
\author{D. Parker}
\affiliation{Department of Physics, University of California, Berkeley, California 94720, USA}
\author{Adolfo G. Grushin}
\affiliation{Department of Physics, University of California, Berkeley, California 94720, USA}
\affiliation{Institut N\'{e}el, CNRS and Universit\'{e} Grenoble Alpes, F-38042 Grenoble, France}
\author{N. L. Nair}
\affiliation{Department of Physics, University of California, Berkeley, California 94720, USA}
\author{J. G. Analytis}
\affiliation{Department of Physics, University of California, Berkeley, California 94720, USA}
\affiliation{Materials Science Division, Lawrence Berkeley National Laboratory, Berkeley, California 94720, USA}
\author{J. E. Moore}
\affiliation{Department of Physics, University of California, Berkeley, California 94720, USA}
\affiliation{Materials Science Division, Lawrence Berkeley National Laboratory, Berkeley, California 94720, USA}
\author{J. Orenstein}
\email{jworenstein@lbl.gov}
\affiliation{Department of Physics, University of California, Berkeley, California 94720, USA}
\affiliation{Materials Science Division, Lawrence Berkeley National Laboratory, Berkeley, California 94720, USA}
\author{Darius H. Torchinsky}
\affiliation{Department of Physics, Temple University, Philadelphia, Pennsylvania 19122, USA}
\affiliation{Temple Materials Institute, Temple University, Philadelphia, PA 19122, USA}
 \date{\today}

\maketitle

\textbf{While all media can exhibit first-order conductivity describing current linearly proportional to electric field, $E$, the second-order conductivity, $\sigma^{(2)}$, relating current to $E^2$, is nonzero only when inversion symmetry is broken. Second-order nonlinear optical responses are powerful tools in basic research, as probes of symmetry breaking \cite{shen1986surface,shen1989surface,dahn1996symmetry,petersen2006nonlinear,harter2017parity}, and in optical technology as the basis for generating currents from far-infrared \cite{robinsonFIR} to X-ray wavelengths \cite{chang1997generation}. The recent surge of interest in Weyl semimetals with acentric crystal structures has led to the discovery of a host of $\sigma^{(2)}$-related phenomena in this class of materials, such as polarization-selective conversion of light to dc current (photogalvanic effects)~\cite{ma2017direct,sun2017circular,osterhoudt2017colossal,ji2018spatially} and the observation of giant second-harmonic generation (SHG) efficiency in TaAs at photon energy 1.5 eV~\cite{wu2017giant,li2018}. Here, we present measurements of the SHG spectrum of TaAs revealing that the response at 1.5 eV corresponds to the high-energy tail of a resonance at 0.7 eV, at which point the second harmonic conductivity is approximately 200 times larger than seen in the standard candle nonlinear crystal, GaAs. This remarkably large SHG response provokes the question of ultimate limits on $\sigma^{(2)}$, which we address by a new theorem relating frequency-integrated nonlinear response functions to the third cumulant (or ``skewness'') of the polarization distribution function in the ground state. This theorem provides considerable insight into the factors that lead to the largest possible second-order nonlinear response, specifically showing that the spectral weight is unbounded and potentially divergent when the possibility of next-neighbor hopping is included.}\\

Second-harmonic generation (SHG) is a subset of $\sigma^{(2)}$-related phenomena in which light of frequency $2\omega$ is generated by excitation at frequency $\omega$. Figure 1 shows a schematic of the apparatus used to measure the spectrum of the SHG response tensor, which is formally written as $\sigma_{ijk}(2\omega;\omega,\omega)$ and hereafter shortened to $\sigma^{\text{shg}}_{ijk}(\omega)$, or simply $\sigma_{ijk}$. The measurements are enabled by a laser/optical parametric amplifier system that generates pulses of duration $\sim$50 fs at a 5 kHz repetition rate in the range of photon energies 0.5 - 1.5 eV.  As the crystals are opaque (optical penetration depth of approximately 200 nm \cite{xuEllipsometry}) the SHG intensity is measured in reflection. As shown in the Figure 1a., a polarizer $P$, and a halfwave plate in front of the sample select the orientation of the electric field (referred to by $\hat{\mathbf{{e}}}$) of the fundamental light at frequency $\omega$, and a second analyzing polarizer $A$ selects the polarization of the light at frequency $2\omega$ that reaches the detector.

\begin{figure}[htp]
	\includegraphics[width=.45\textwidth]{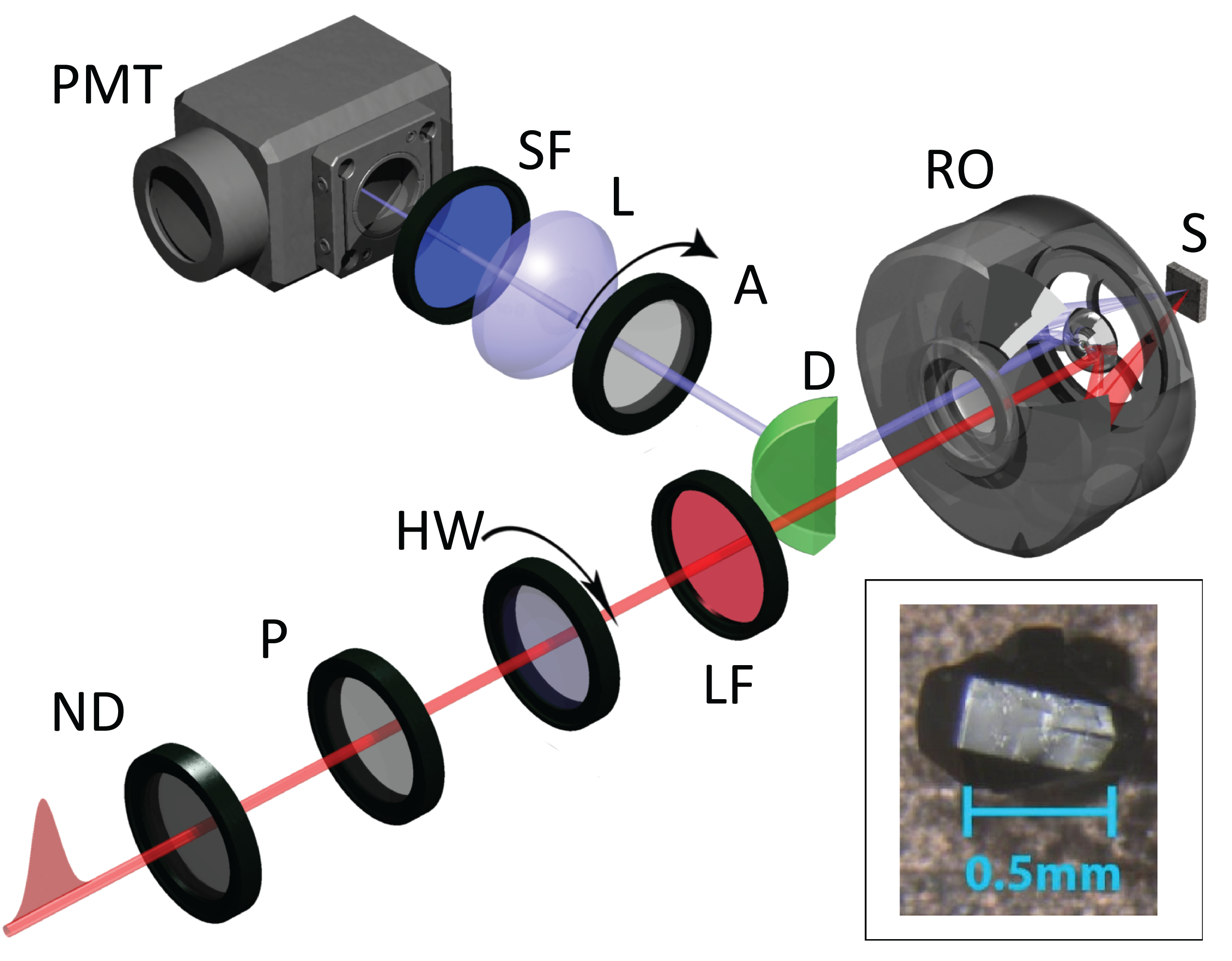}
\caption{
\textbf{Schematic of the experimental apparatus used to study the spectrum of the nonlinear optical response
in TaAs.} Laser pulses of pulse width 50 fs, at a repetition rate of 5 kHz with photon energy between 0.5 eV - 1.6 eV are generated by an optical parametric amplifier coupled to an regeneratively amplified laser. The incident radiation passes through a neutral density filter (ND), polarizer ($P$), and a low pass filter (LF), to lower the incident intensity, and to remove spurious polarizations, and  wavelengths, respectively. A halfwave plate (HW) is used to rotate the incident linear polarization over 360$^o$. The incident laser radiation is then focused on to the TaAs sample (S) using a reflective objective (RO), which is insensitive to the photon energy, in contrast with conventional refractive microscope objectives. The SHG radiation from the sample is collimated by the RO, and picked off by a D-mirror (D), and passed through an analyzing polarizer ($A$), whose polarization state is determined by the scan being measured (see text). The radiation then passes through a stack of short pass filters (SF) to remove any incident light, and then focused by a lens (L) onto a photomultiplier tube (PMT) for detection. Inset: A single crystal of TaAs with the $(112)$ facet visible.}
	\label{schematic}
\end{figure}

The structure of TaAs (point group $4mm$) contains two perpendicular glide planes, which are equivalent to mirror planes for optical response functions \cite{TaAsWeng,li2016weyl}. The normal directions of these two planes determine the two equivalent directions of the tetragonal unit cell, which we label as $x$ and $y$. The direction perpendicular to the $xy$ plane is the polar, or $z$ axis.  The $4mm$ point group allows three independent nonvanishing components of the conductivity tensor, $\sigma_{zzz}$, $\sigma_{zxx}=\sigma_{zyy}$, and $\sigma_{xzx}=\sigma_{yzy}$. To avoid using a surface obtained by cutting and subsequent polishing, which are known to affect the nonlinear optical response, we perform nonlinear reflectance measurements using the (112) surface, as it forms a naturally flat and smooth growth facet. Figure \ref{schematic}b shows a photograph of the (112) surface used for the measurement. The two high symmetry directions in the plane of this surface are [1,-1,0], which is perpendicular to the polar axis, and [1,1,-1]. 

We performed two pairs of polarization scans at each wavelength. In the first pair, $\hat{\mathbf{{e}}}$ and $A$ are co-rotated with their relative angle fixed at 0 and 90$^{\circ}$. 
This is equivalent to rotating the sample, but with the advantage that the position of the laser focus on the sample surface remains constant. These scans are primarily used to determine the high symmetry directions. However, as will be discussed below, they provide additional evidence for the existence of a sharp resonance in the nonlinear response function in the infrared. In the second pair of scans $A$ is fixed parallel to either of the [1,-1,0] or [1,1,-1] high symmetry directions and $\hat{\mathbf{{e}}}$ is rotated over 360$^{\circ}$. From this pair of scans we obtain the three combinations of tensor components that are available from reflection measurements performed on the (112) surface: $|\sigma_{zxx}|$, $|\sigma_{xzx}|$, and $|\sigma_{\mathrm{eff}}|\equiv  |\sigma_{zzz}+2\sigma_{zxx}+4\sigma_{xzx}|$ \cite{seeApp}. Analysis of the polarization scans at a given wavelength determines the relative magnitudes of these tensor components, i.e., $|\sigma_{zxx}|/|\sigma_{\mathrm{eff}}|$ and $|\sigma_{xzx}|/|\sigma_{\mathrm{eff}}|$.

Measurements of absolute, as opposed to relative, magnitudes of $\sigma_{ijk}$ components over a broad range of frequency were accomplished by using GaAs to calibrate the response at 1.5 eV. Determining the effective $\sigma_{ijk}(\omega)$ at lower photon energies required characterization of the wavelength dependence of all components of the optical setup, such as wavelength-sorting filters, attenuators, and photodetectors, as well as spectral variation of the laser focal spot size and pulse duration. An extensive discussion of the calibration procedure is provided in the Supplementary Information.

Polarization scans were performed at incident wavelengths in the interval from 800 to 2500 nm and a complete library of these data is included in the Supplementary Information. Figure \ref{polar_all}(a-c) shows representative scans at wavelengths 800, 1560, and 2200 nm (1.55, 0.80, and 0.56 eV, respectively) with co-rotating parallel polarizations. The co-rotation plot at 800 nm illustrates the extreme anisotropy of the nonlinear optical response of TaAs. The solid curve through the data points is proportional to $|\sigma_{zzz}|^2\cos^6\theta$, which is the dependence predicted for a crystal that responds only to $E$ parallel to its polar axis and radiates second harmonic light that is likewise fully $z$-polarized. The polar plots at longer wavelength indicate that the amplitudes of the off-diagonal components, $|\sigma_{zxx}|$ and $|\sigma_{xzx}|$, begin to increase relative to $|\sigma_{\text{eff}}|$ as the fundamental photon energy approaches 0.7 eV from above, although they remain approximately an order of magnitude smaller, as is shown below.  

\begin{figure*}[htp]
\includegraphics[width=0.99\textwidth]{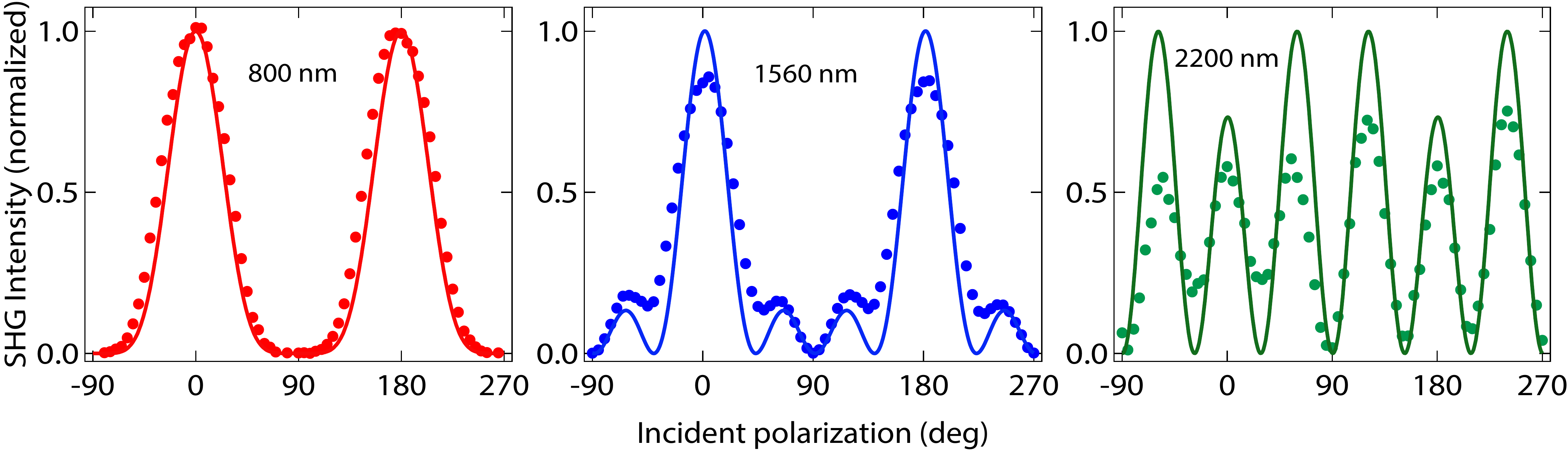}
	\caption{
	\textbf{Second harmonic generation polarimetry.} SHG intensity measured in the ``parallel'' scan as a function of the incident polarization angle, plotted on a normalized scale, for three different incident photon wavelengths. The polar pattern changes from a two-fold pattern at shorter wavelengths (800 nm=1.55 eV) to a six-fold pattern at longer wavelengths (2200 nm=0.56 eV)}
	\label{polar_all}
\end{figure*}

\begin{figure}[htp]
		\includegraphics[width=0.475\textwidth]{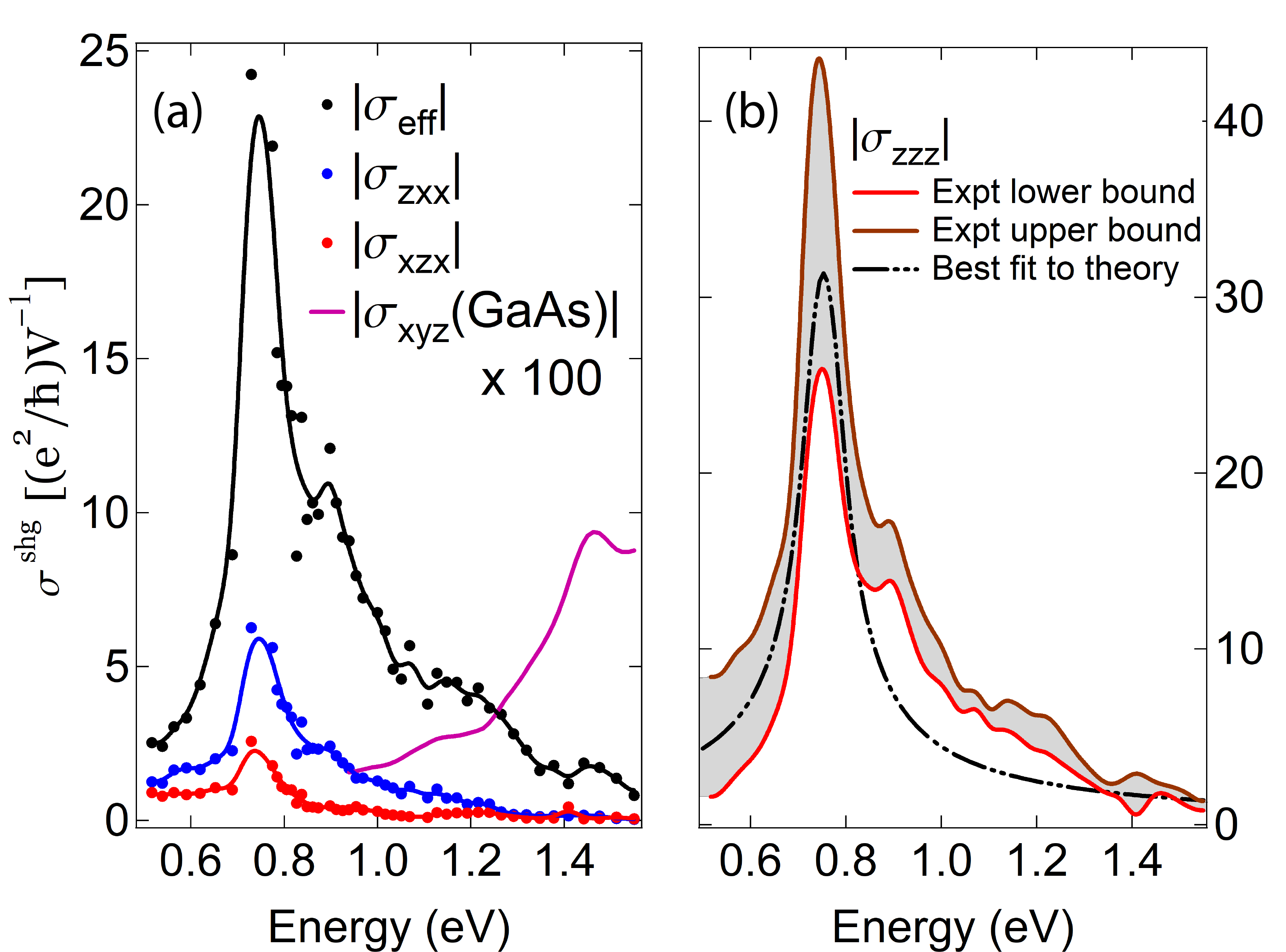}
		\caption{
	\textbf{Measured nonlinear optical conductivity as a function of incident photon energy.} (a) Spectra of the conductivity components $|\sigma_{zxx}|$, $|\sigma_{xzx}|$, and $|\sigma_{\text{eff}}|=|\sigma_{zzz}+4\sigma_{xzx}+2\sigma_{zxx}|$. The solid lines are guides to the eye. The nonlinear optical conductivity of GaAs, $|\sigma_{xyz}|$ is multiplied by 100 and plotted for comparison. (b) The (expt) lower and upper bounds of $|\sigma_{zzz}|$ are estimated using measured values of $|\sigma_{\text{eff}}|$, $|\sigma_{zxx}|$, and $|\sigma_{xzx}|$ from experiment. The dashed black line depicts the best fit to $|\sigma_{zzz}|$ using the phenomenological model described in the text.}
	\label{spectra}
\end{figure}

Figure \ref{spectra} illustrates the spectral dependence of the SHG response function of a TaAs crystal at room temperature. Figure \ref{spectra}a shows $|\sigma_{\mathrm{eff}}|$, $|\sigma_{zxx}|$, and $|\sigma_{xzx}|$ vs. fundamental photon energy as solid circles (the solid lines are guides to the eye). The $y$-axis scale expresses $\sigma^{\text{shg}}$ in units of the conductance quantum per Volt, to facilitate comparison with theory. In SI units, the corresponding peak value of $\sigma^{\text{shg}}$ at 0.7 eV is approximately $5\times10^{-3}\;(\Omega\text{ V})^{-1}$. The dashed line is the spectrum of $\sigma_{xyz}$ in GaAs as reported in Ref. \cite{bergfeld2003second}, multiplied by a factor 10 so that it can be compared with the TaAs spectra. As reported previously, the SH response measured at 1.5 eV in TaAs exceeds the peak response of GaAs by a factor $\sim{}10$. The measurements reported here show that $|\sigma_{\mathrm{eff}}|$ becomes even larger with decreasing frequency, reaching a peak near 0.7 eV where it is $\sim 2\times 10^2$ larger than the maximum response of GaAs.

Although the diagonal response function $|\sigma_{zzz}|$ is not determined directly from the SHG intensity, its range of values is highly constrained through analysis of the polarization scans \cite{seeApp}. The shaded region in Figure \ref{spectra}b illustrates the upper and lower bounds on $|\sigma_{zzz}|$. The dashed line that lies within the shaded region is a best fit to a phenomenological model described below, which is seen to capture the basic features of the resonance enhancement of $|\sigma_{zzz}|$.\\

%
%

The key features of the nonlinear conductivity in TaAs --– strong anisotropy and a single, sharp resonant peak --– suggest a description in terms of a one-dimensional (1D) two-band model, modified for a 3D system.  Recently it was shown that the Rice-Mele (RM) tight-binding model of a polar semiconductor~\cite{RM82} provides an excellent description of the spectrum of the shift current response in single-layer monochalcogenide semiconductors \cite{rangel2017large}. As we discuss below, for such models there is precise correspondence between shift current and SHG spectra \cite{MN16}.  Moreover, the energies probed in our experiment are much larger than either the Fermi energy \cite{xu2015discovery,TaAsLv} or plasma frequency \cite{xu2016optical} of TaAs, suggesting that the semimetallic part of its bandstructure is not relevant to the nonlinear response at photon energies of order $1$ eV. 

\begin{figure}[htp]

	\includegraphics[width=0.45\textwidth]{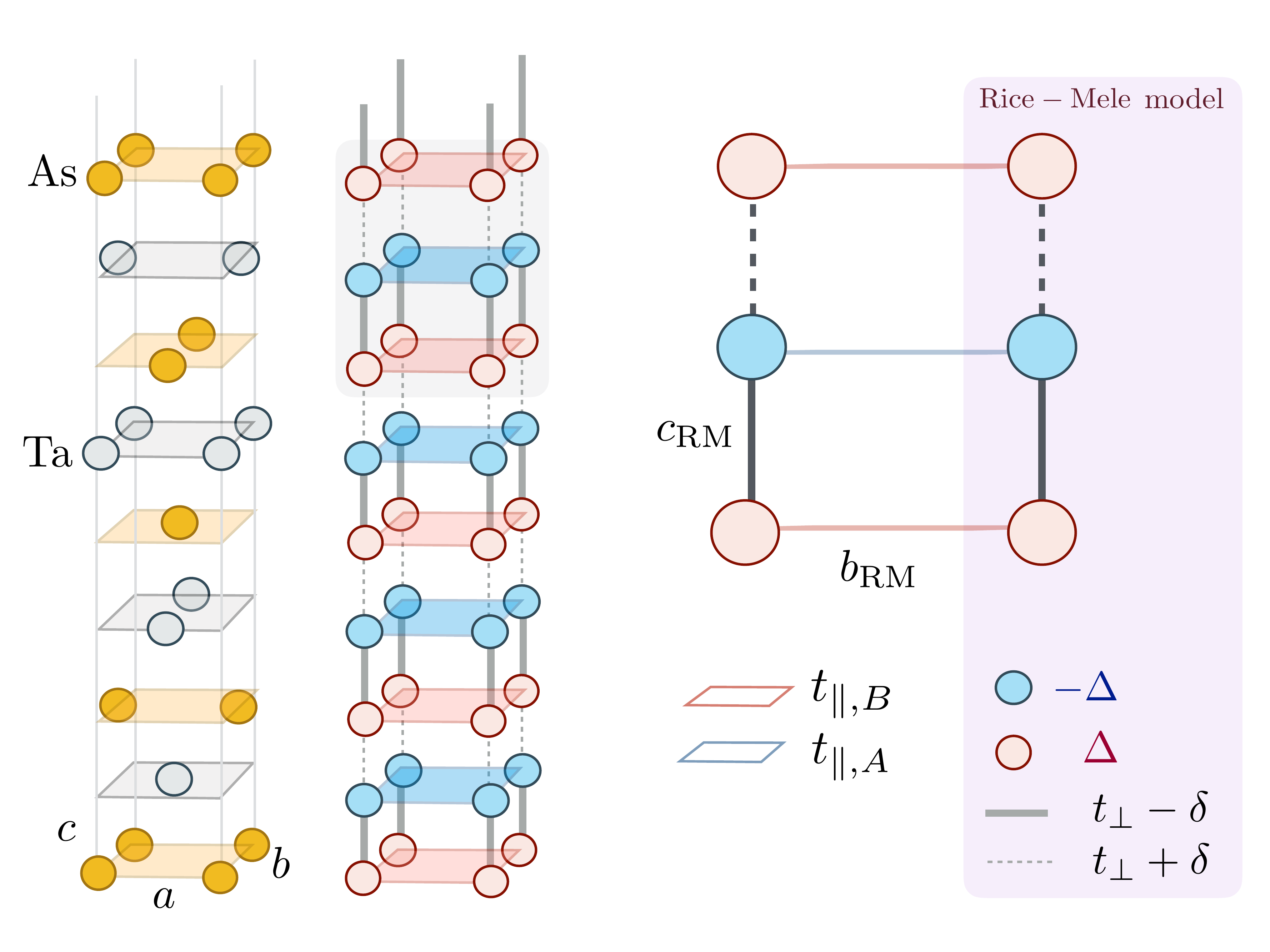}

	\caption{
	\textbf{The Rice-Mele model.} Depiction of the model of coupled Rice-Mele chains (center) used to reproduce the experimentally observed second harmonic generation in TaAs (left).
A zoom on the shaded unit-cell is depicted on the right specifying the different hopping terms present in the model. 
}
	\label{Fig4}
\end{figure}

These considerations lead us to compare the measured SHG spectra with the predictions for a system described by the RM Hamiltonian,
\begin{equation}
\abovedisplayskip=5pt
H_{RM}=t\cos(k_za/2)\sigma_z+\delta\sin(k_z a/2)\sigma_y+\Delta \sigma_z,
\label{fig:RM1D}
\belowdisplayskip=5pt
\end{equation}
where the $\sigma_i$ are the Pauli matrices. As shown in Figure \ref{Fig4}, $H_{RM}$ describes a 1D chain of atoms along the $z$ direction, in which inversion symmetry is broken by alternating on-site energies ($\pm \Delta$) and hopping amplitudes ($t\pm \delta$). Despite its relative simplicity, the RM model has played a key role in the development of the modern theory of polarization~\cite{resta2007theory} and recently in strategies to enhance nonlinear optical response functions \cite{tan2017upper}.



The optical response of the RM Hamiltonian captures qualitative features of the nonlinear optical response in TaAs, as $\sigma^{\text{shg}}_{ijk}(\omega)$ is polarized parallel to the polar axis and exhibits a sharp peak at the threshold for absorption. However, to allow a quantitative comparison with experiment the RM model must be modified to include an interchain coupling to smooth the 1D van Hove singularity that is obtained for an isolated chain.


To extend the model to 3D, we consider an array of RM chains parallel to $z$ and allow electrons to hop laterally in the $A$ and $B$ atom layers (see Figure \ref{Fig4}). The interchain hopping adds an additional term $t_{AB}(\cos k_x a_x+\cos k_y a_y)$ to the coefficient of $\sigma_z$ in $H_{RM}$, where $t_{AB}\equiv t_{\parallel,A}-t_{\parallel,B}$. The shift current spectrum in this model can be calculated following \cite{SipeShkrebtii} (see \cite{seeApp}),
\begin{equation}
\sigma_{zzz}^{\text{shift}}(\omega)
= \frac{2e^3}{\hbar\Delta}
\left( \frac{1}{4\pi} \right)^3 \frac{c^2}{ab} 
F(\tilde{\omega}; \tilde{\delta}, \tilde{t}, \tilde{t}_{AB}),
\end{equation}
where $F$ is a dimensionless function of the parameters of $H_{RM}$, and of frequency $\omega$ (the tildes indicate normalization by $\Delta$); $a$, $b$, and $c$ are lattice constants in the $x$, $y$, and $z$ directions, respectively \cite{seeApp}. The real part of the SHG conductivity can be expressed in terms of the shift conductivity as follows \cite{MN16}, 
\begin{equation}
\text{Re}\left\lbrace \sigma^{\text{shg}}_{zzz}(\omega)\right\rbrace  =-\frac{1}{2} \sigma_{zzz}^{\text{shift}}(\omega) + \sigma_{zzz}^{\text{shift}}(2\omega)
\label{1-2_photon}
\end{equation}
Equation (\ref{1-2_photon}) shows that the SHG response is the sum of two terms that are opposite in sign but have the same functional form when the frequency argument is scaled by a factor of two. The first (``one-photon'') term corresponds to the resonant condition $\hbar\omega=E_{\text{gap}}(k)$ and the second (``two-photon'') term corresponds to $\hbar\omega=E_{\text{gap}}(\textbf{k})/2$, where $E_{\text{gap}}(\textbf{k})$ is the energy gap at wavevector $\textbf{k}$. Clearly the total $\sigma^{\text{shg}}_{zzz}$ response integrates to zero, but we shall show momentarily that for some models each term has a geometrical interpretation in terms of the skewness or intrinsic asymmetry of the ground-state polarization distribution.

The dash-dotted curve in Figure \ref{spectra}b is a fit to $|\sigma^{\text{shg}}_{zzz}|$ based on the two-photon term in equation (\ref{1-2_photon}). 
Given the simplicity of the model relative to the complexity of the full 3D bandstructure of TaAs, the 3D version of the RM model describes the data remarkably well. The peak position, low, and high energy tails of the spectrum are best described using parameters $\tilde{t}=1.5$, $\tilde{\delta}=1.4$, $\tilde{t}_{AB}=0.02$, and $\Delta=0.428$. The lattice parameters for TaAs are $c=1.165$ nm, and $b=a=0.344$ nm.

The theory discussed above helps to identify the factors that contribute to a large SHG response. The first is a polar crystal structure such as $4mm$ (TaAs, NbAs, etc.) or $mm2$ (single layer monochalcogenides) consistent with a phenomenological description in terms of coupled ferroelectric chains.  However, even when this description is possible there is a global bound on the frequency integrated second-order response, $\Sigma_z^{\text{shift}}$, of a system described by $H_{RM}$ that is given by,
\begin{equation}
\Sigma_z^{\text{shift}}\equiv\int\sigma_{zzz}^{\text{shift}}(\omega)d\omega=\frac{e^3}{\hbar^2}\frac{c^2}{ab}G(\tilde{t},\tilde{\delta}).
\label{bigSigma}
\end{equation}
where $G$ is also a dimensionless function of the dimensionless parameters, with a global maximum 0.604 \cite{seeApp}. Given this bound on $\Sigma_z^{\text{shift}}$, it is clear that the single most important factor in optimizing the second-order response of a 3D crystal is the effective aspect ratio parameter $c^2/ab$. 

We conclude by addressing the question of ultimate bounds on $\Sigma^{\text{shift}}_z$, which is relevant not only to SHG, but to sensitivity of photodetectors and efficiency of solar cells based on Berry curvature generated intrinsic photogalvanic effects \cite{MN16,MooreOrenstein} as well. We have discovered a new connection between $\Sigma^{\text{shift}}_z$ and the modern theory of polarization that generalizes the formulae for spectral weight beyond the RM Hamiltonian. Specifically we show that $\Sigma^{\text{shift}}_z$ is unbounded and potentially divergent when the possibility of next-neighbor hopping is included.

As demonstrated in the Supplementary Information, $\Sigma_z^{\text{shift}}$ in two band systems, regardless of dimensionality $d$ and range of electron hopping amplitude is given by,

\begin{equation}
\label{sumrule}
\Sigma_z^{\text{shift}}=\frac{2\pi e^3}{V\hbar^2}C_3,
\end{equation}
where,
\begin{equation}
C_3 \ =\ -\frac{V}{(2\pi)^d} \int d^dk \; \text{Im}\left[ c_3 - 3 c_2c_1 + 2 c^3_1 \right],
\end{equation}
$c_n \equiv \bra{u_0(k)} (i \partial_k)^n \ket{u_0(k)}$, and the periodic gauge is assumed for the valence-band Bloch wavefunction $\psi_{0k}(r) = u_{0k}(r) e^{i k r}$. 
The quantity $C_3$ is a member of a set of gauge invariant quantities, $C_n$, that are cumulants of the electronic polarization \cite{hetenyi_cumulants,souza2000polarization}. The quantity $C_1$ is exactly the average macroscopic polarization, which coincides with the first moment of the polarization distribution \cite{resta2007theory}.  Accordingly, $C_3$ is the third cumulant, or ``skewness'' of the distribution, which vanishes in the presence of inversion symmetry. $C_1$ is controlled by the center-of-mass location of the polarization, while $C_3$ describes the intrinsic asymmetry in the shape of the polarization distribution, independent of its center-of-mass location.

In addition to providing a very satisfying connection between the nonlinear response and ground state fluctuations of the polarization, equation (\ref{sumrule}) speeds the search for effective Hamiltonians with large nonlinear response. For instance, we find that the bound on $\Sigma_z^{\text{shift}}$ that is obtained in the 3D RM model is exceeded when 
 next-neighbor hopping is introduced, and can diverge when the fundamental gap is driven to zero. This result is consistent with the recent analysis of Tan and Rappe \cite{tan2017upper}, who suggested that $\Sigma_z^{\text{shift}}$ can be enhanced when the range of inter-site hopping becomes larger than the lattice parameter.

In this work we reported SHG spectra of TaAs demonstrating a ten-fold resonant enhancement of the response at 0.7 eV as compared with the previously reported, and already quite large, response at 1.5 eV. We showed that the spectral dependence and absolute magnitude of the SHG response function is well captured by a model based on coupled ferroelectric Rice-Mele chains, suggesting that the giant nonlinear response in TaAs is related to its strong polar nature. The relationship between the RM phenomenology and the actual electronic wavefunctions and bandstructure of TaAs remains to be understood. Finally, we presented a new theorem linking the spectral weight of the nonlinear conductivity to the third gauge-invariant cumulant of the ground state. The cumulant $C_3$ is likely to be more computationally tractable, and we anticipate that this theorem will facilitate algorithmic searches for materials with large nonlinear response functions.

\section{Acknowledgements}
Measurements and modeling were performed at the Lawrence Berkeley National Laboratory in the Quantum Materials program supported by the Director, Office of Science, Office of Basic Energy Sciences, Materials Sciences and Engineering Division, of the U.S. Department of Energy under Contract No. DE-AC02-05CH11231. Spectroscopy measurements were performed at the Department of Physics, Temple University. The authors would like to thank B. Xu for sharing ellipsometry data of TaAs. We would also like to thank Balasz Het\'enyi for helpful communication. 
J.O. and L.W. received support for performing and analyzing optical measurements from the Gordon and Betty Moore Foundation's EPiQS Initiative through Grant GBMF4537 to J.O. at UC Berkeley.  Sample synthesis by N.N. and J.A. was supported by the Gordon and Betty Moore Foundation’s EPiQS Initiative through Grant GBMF4374. Work by N.N. and J.A. was supported by the Office of Naval Research under the Electrical Sensors and Network Research Division, Award No. N00014-15-1-2674. T.M. and J.E.M. were supported by the Quantum Materials program under the U.S. Department of Energy. T.M was supported by the Gordon and Betty Moore Foundation's EPiQS Initiative Theory Center Grant to UC Berkeley. A.G.G. was supported by the Marie Curie Program under EC Grant agreement No. 653846. J.E.M. received support for travel from the Simons Foundation. D.H.T. acknowledges startup funds from Temple University. D.P. received support from the NSF GRFP,  DGE 1752814. The authors would like to thank Nobumichi Tamura for his help in performing crystal diffraction and orientation on beamline 12.3.2 at the Advanced Light Source. N. Tamura and the ALS were supported by the Director, Office of Science, Office of Basic Energy Sciences, of the U.S. Department of Energy under Contract No. DE-AC02-05CH11231.

\textit{Competing Interests: } The authors declare that they have no
competing financial interests.

\textit{Correspondence: } Correspondence and requests for materials
should be addressed to D.H.T. (dtorchin@temple.edu) and J.O. (jworenstein@lbl.gov).

\section{Author Contribution}
J.O., L.W., S.P., and D.H.T. conceived the project. D.H.T., S.P., L.W., B.L., M.R., and J.D.T. performed the SHG spectroscopy and calibration measurements. J.O., S.P., L.W., and D.H.T. analyzed the data. J.O., A.G.G., T.M., D.P., and J.E.M. developed the one-dimensional model and calculated theoretical parameters. N.N. and J.A. grew the crystals and characterized the crystal structure.  J.O., S.P., L.W., D.P., T.M., A.G., and D.H.T. prepared the manuscript with assistance from all authors.

\appendix


\section{Data acquisition and processing}
\subsection{Extracting susceptibility components from SHG polar patterns}
To calculate the components of the third rank susceptibility tensor $\overleftrightarrow{\boldsymbol{\chi}}^{(2)}$ from the measured data, we start with the relation defining the second harmonic susceptibility in terms of the electric fields of the incident $\mathbf{E}^{1\omega}$ and radiated $\mathbf{E}^{2\omega}$ fields:
\begin{equation}
E^{2\omega}_i=\chi_{ijk}^{\text{shg}}E^{1\omega}_jE^{1\omega}_k,
\label{eq:chi_definition}
\end{equation}
where a sum over repeated indices is implied. We use the notation $\hat{x},\hat{y},\hat{z}$, to correspond to the crystalline axes $[1,0,0]$, $[0,1,0]$, $[0,0,1]$, and for brevity, we shall drop the superscript ``shg'' in the following. The crystal structure of TaAs corresponds to the point group $4mm$, which allows only three distinct components of the $\chi_{ijk}$ tensor: $\chi_{zzz}$; $\chi_{zxx}=\chi_{zyy}$; $\chi_{xxz}=\chi_{xzx}=\chi_{yyz}=\chi_{yzy}$. As noted in the main text, the experiments were performed by measuring the second harmonic intensity with light incident on the naturally occuring $(112)$ facet of a TaAs single crystal. The SHG intensity was measured in three polarization ``channels'':
\begin{enumerate}
\item \textit{parallel}: the polarization state of the analyzer is set to be parallel to the polarization state of the incident radiation, while both rotate from $0^\circ$ to $360^\circ$
\item \textit{vertical}: the polarization state of the analyzer is set to be parallel to the in-plane polar direction $(1,1,-1)$ of the TaAs crystal, while the polarization state of the incident radiation rotates from $0^\circ$ to $360^\circ$
\item \textit{horizontal}: the polarization state of the analyzer is set to be perpendicular to the in-plane polar direction $(1,1,-1)$ of the TaAs crystal, while the polarization state of the incident radiation rotates from $0^\circ$ to $360^\circ$
\end{enumerate}

It is useful to label the in-plane component of the polar axis $[1,1,-1]$ as the axis ``$\gamma$'', and correspondingly label the perpendicular in-plane direction $[-1,1,0]$ as the axis ``$\alpha$''. 
 If the incident light has intensity $I_0$, with linearly polarized electric field expressed by $\mathbf{E}^{1\omega}=\sqrt{I_0}\left(\hat{\alpha}\sin\theta+\hat{\gamma}\cos\theta\right)$, in the parallel channel, the intensity can be expressed in terms of the angle $\theta$ of the incident polarization state with respect to the ``$\gamma$'' axis:

\begin{align*}
I_{\text{para}}(\theta)
&=\left|E^{2\omega}_{\alpha}\sin\theta+E^{2\omega}_{\gamma}\cos\theta\right|^2\\
&=\left|\frac{1}{\sqrt{2}}\left(-E^{2\omega}_x+E^{2\omega}_y\right)\sin\theta+\frac{1}{\sqrt{3}}\left({E^{2\omega}_x+E^{2\omega}_y-E^{2\omega}_z}\right)\cos\theta\right|^2.
\end{align*}


We wish to express this in terms of the tunable quantities $E^{1\omega}_z=-\tfrac{E^{1\omega}_{\gamma}}{\sqrt{3}}$ and $E^{1\omega}_{\gamma}=\sqrt{I_0}\cos\theta$. To do this, we first expand $\v{E}^{2\omega}$ in terms of $\v{E}^{1\omega}$ via
\begin{equation}
	 E^{2\omega}_x=2\chi_{xxz}E^{1\omega}_xE^{1\omega}_z,
	 \qquad E^{2\omega}_y=2\chi_{yyz}E^{1\omega}_yE^{1\omega}_z,
	 \qquad E^{2\omega}_z=\chi_{zxx}\left(E^{1\omega}_x\right)^2+\chi_{zyy}\left(E^{1\omega}_y\right)^2+\chi_{zzz}\left(E^{1\omega}_z\right)^2,
\end{equation}
and transform back from the $xyz$ to the $\alpha\gamma$ basis via
\begin{equation}
	E^{1\omega}_x=-\frac{E^{1\omega}_{\alpha}}{\sqrt{2}}+\frac{E^{1\omega}_{\gamma}}{\sqrt{3}},\qquad
	E^{1\omega}_y=\frac{E^{1\omega}_{\alpha}}{\sqrt{2}}+\frac{E^{1\omega}_{\gamma}}{\sqrt{3}},\qquad
	E^{1\omega}_z=-\frac{E^{1\omega}_{\gamma}}{\sqrt{3}}.
\end{equation}
This yields


\begin{align*}
I_{\text{para}}(\theta)
&=\frac{I_0^2}{3}\left|\chi_{xxz}\cos\theta\sin^2\theta+\left(\frac{4}{3}\chi_{xxz}\cos^2\theta+\chi_{zxx}\left(\sin^2\theta+\frac{2}{3}\cos^2\theta\right)+\frac{1}{3}\chi_{zzz}\cos^2\theta\right)\cos\theta\right|^2\\ 
&=\frac{I_0^2}{27}\left|3\chi_{xxz}\cos\theta\sin^2\theta+\left(4\chi_{xxz}\cos^2\theta+\chi_{zxx}\left(3\sin^2\theta+2\cos^2\theta\right)+\chi_{zzz}\cos^2\theta\right)\cos\theta\right|^2\\ 
&=\frac{I_0^2}{27}\left|(6\chi_{xxz}+3\chi_{zxx})\cos\theta\sin^2\theta+\left(4\chi_{xxz}+2\chi_{zxx}+\chi_{zzz}\right)\cos^3\theta\right|^2.
\end{align*}
We define $\chi_{\text{eff}}=\chi_{zzz}+2\chi_{zxx}+4\chi_{xxz}$, which gives us the expression,
\begin{equation}
I_{\text{para}}=\frac{I_0^2}{27}\left|3(2\chi_{xxz}+\chi_{zxx})\cos\theta\sin^2\theta+\chi_{\text{eff}}\cos^3\theta\right|^2.
\label{eq:para}
\end{equation}
We can similarly obtain expressions for the other two channels:
\begin{align}
I_{\text{vertical}}(\theta)
=\left|E^{2\omega}_3\right|^2
=\left|\frac{1}{\sqrt{3}}\left({E^{2\omega}_x+E^{2\omega}_y-E^{2\omega}_z}\right)\right|^2
=\frac{1}{27}\left(3\chi_{zxx}\sin^2\theta+\chi_{\text{eff}}\cos^2\theta\right),
\label{eq:vert}
\end{align}
and
\begin{align}
I_{\text{horiz}}(\theta)
=\left|E^{2\omega}_{\alpha}\right|^2
=\left|\frac{1}{\sqrt{2}}\left(-E^{2\omega}_x+E^{2\omega}_y\right)\right|^2
=\frac{1}{3}I_0^2|\chi_{xxz}|^2\;\sin2\theta.
\label{eq:horiz}
\end{align}

\begin{figure}\centering
\includegraphics[scale=0.35]{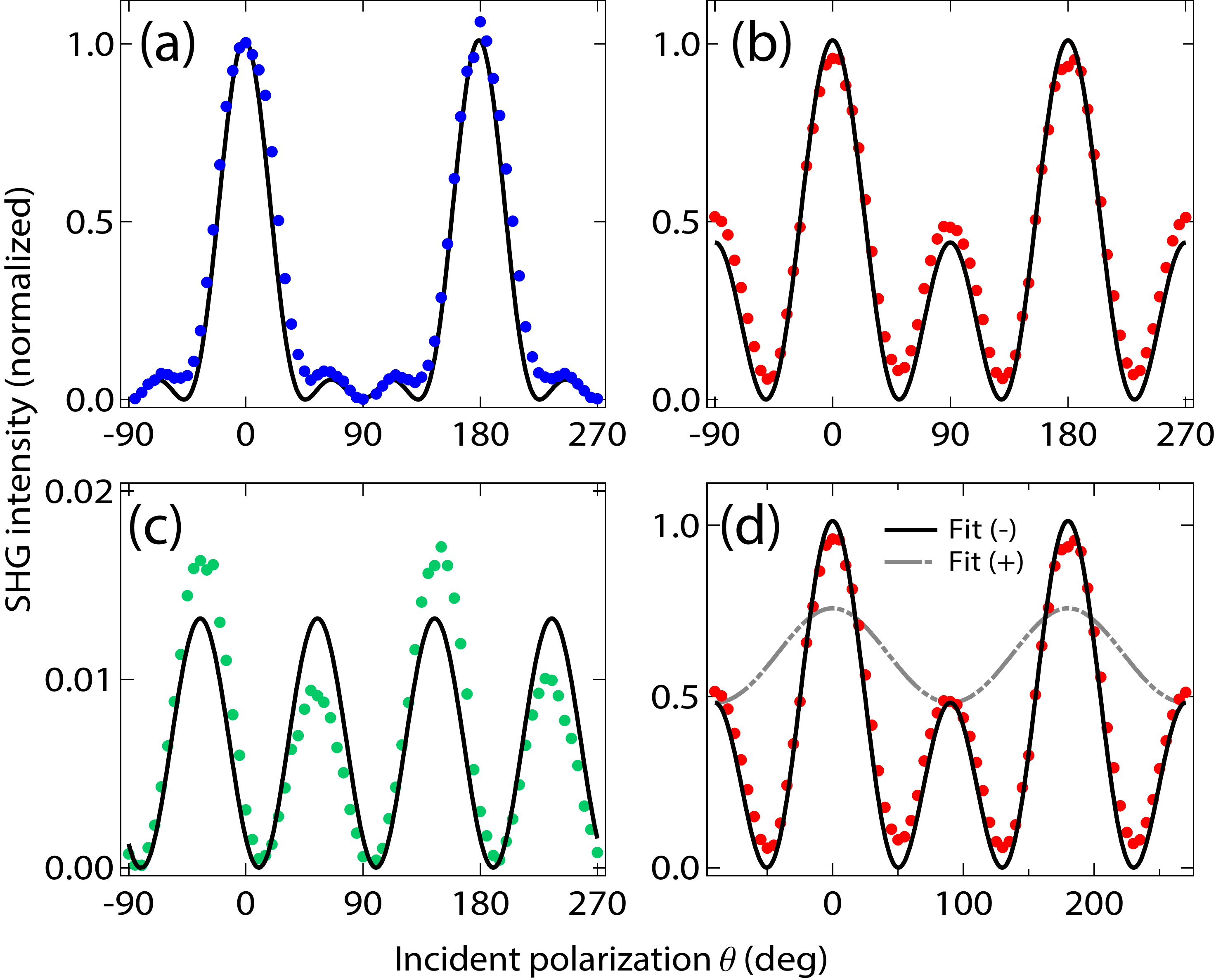}
\caption{Minimized mean-squared deviation fits (solid lines) to the SHG intensity data (points) as a function of the incident polarization angle $\theta$, in three polarization channels (a) parallel, (b) vertical, (c) horizontal, for incident photon wavelength $\lambda=1400$nm, plotted on a normalized scale. (d) Fits to the SHG intensity in the vertical channel assuming the relative sign between $\chi_{zxx}$ and $\chi_{zzz}$ to be $(+)$ (dashed line), and $(-)$ (solid line).}
\label{polar_fits}
\end{figure}

Figure~\ref{polar_fits}d shows the SHG intensity data as a function of the polarization of incident photons in the parallel, vertical, and horizontal channel, with fits to the expressions in equation \eqref{eq:para}, \eqref{eq:vert}, \eqref{eq:horiz} respectively. The amplitudes of the fitting parameters $|\chi_{xxz}|, |\chi_{zxx}|, |\chi_{\text{eff}}|$ thus obtained are then multiplied by various wavelength-dependent correction factors described below.

Although the fitting scheme described above is not sufficient to obtain the complete amplitude and phase information for the three components $|\chi_{xxz}|, |\chi_{zxx}|, |\chi_{\text{eff}}|$, we are nevertheless able to definitively say that for all incident photon energies, $\chi_{zxx}$ and $\chi_{zzz}$ have a relative phase of $\pi$. Figure~\ref{polar_fits} shows the best fits to the data at incident photon wavelength $\lambda=\SI{1400}{\nano\meter}$ in the ``vertical'' channel, using two contrasting assumptions for the relative signs of the fitting parameters, $(+1)$ or $(-1)$. The fits with relative $(-)$ signs are are observed to be better approximations to the data.\\

With the additional information about the relative signs, the calculated values of $|\chi_{\text{eff}}|=|\chi_{zzz}+2\chi_{zxx}+4\chi_{xxz}|$ can be used to place bounds on the values of $|\chi_{zzz}|$ with an upper bound (UB) and lower bound (LB) expressed as (see figure~3 in the main text):

\begin{align}
|\chi_{zzz}^{UB}| &= |\chi_{\text{eff}}|+2|\chi_{zxx}|+4|\chi_{xxz}| \\
|\chi_{zzz}^{LB}| &= |\chi_{\text{eff}}|+2|\chi_{zxx}|-4|\chi_{xxz}|.
\end{align}

\subsection{Wavelength dependent correction factors}

We provide a brief summary of the factors that must necessarily be taken into consideration to accurately determine the second harmonic susceptibility tensor elements $\chi_{ijk}^{\text{shg}}$ and the procedure we followed to experimentally measure them.

The electric field of the incident laser pulse can be well approximated by a simple Gaussian field as
\begin{equation}\label{eq:field}
\mathbf{E}_{\omega}(\mathbf{r},t) =E_0e^{-i(k_0z-\omega t)}\exp\left(-\frac{x^2+y^2}{w_0^2}\right)\exp{\left[-\left(\frac{z-ct}{\tau}\right)^2 \right]}\hat{x}
\end{equation}
where $E_0$ is the electric field amplitude, $k_0$ is the field wavevector magnitude, $\omega$ is the fundamental light frequency, $c$ is the speed of light, and $w_0$ is the electric field beam waist~\cite{Quesnel1998theory}, here taken as the point from which the SHG field arises. We have assumed that the beam is transversely polarized purely in the $\hat{x}$ direction for convenience. As a consequence of Poynting's Theorem, the relationship between the integrated intensity of the incident, fundamental frequency field $I_\omega$ and the associated electric field can be given by
\begin{equation}
I_\omega  = \frac{1}{2}\epsilon_0\int \left\langle|\mathbf{E}_{\omega}(\mathbf{r},t)|^2\right\rangle d\mathbf{r}
\end{equation}
where the brackets denote time averaging. For the field as given by equation~\eqref{eq:field}, the integration is easily performed to yield
\begin{equation}\label{eq:fundint}
I_\omega \propto E_0^2 w_0 \tau
\end{equation}

The penetration depth of the wavelengths used in this study is relatively short in \ce{TaAs}. Thus we may consider the limit of non-depleted incident fields. The radiated second harmonic electric field is then proportional to the second-order induced polarization as $\mathbf{E}_{2\omega}(\mathbf{r},t) \propto \mathbf{P}_{2\omega}(\mathbf{r},t) = \epsilon_0\overleftrightarrow{\boldsymbol{\chi}}^{\text{shg}}\mathbf{E}_{\omega}(\mathbf{r},t)\mathbf{E}_{\omega}(\mathbf{r},t)$. 
Hence, the intensity of the second harmonic field $I_{2\omega}$ is given by
\begin{equation}
I_{2\omega}  = \frac{1}{2}\epsilon_0\int \left\langle|\mathbf{E}_{2\omega}(\mathbf{r},t)|^2\right\rangle d\mathbf{r}\propto \int \left\langle|\overleftrightarrow{\boldsymbol{\chi}}^{\text{shg}}|^2|\mathbf{E}_{\omega}(\mathbf{r},t)|^4\right\rangle d\mathbf{r}\propto (E_0^2\chi_{ijk}^{\text{shg}})^2 w_0 \tau.
\end{equation}
When this latter result is combined with equation~\eqref{eq:fundint} to eliminate the electric field $E_0$, we determine that
$I_{2\omega}\propto(\chi_{ijk}^{\text{shg}})^2 I_\omega^2/w_0\tau$. However, for Gaussian fields such as in equation~\eqref{eq:field}, the beam waist at the focus $w_0$ can be described in terms of the incident beam diameter $d$, lens focal length $f$ and beam wavelength $\lambda$ as $w_0\propto f\lambda/d$, yielding
\begin{equation}
I_{2\omega}\propto (\chi_{ijk}^{\text{shg}})^2\frac{I_\omega^2 d}{f\lambda\tau}.
\end{equation}
Therefore, accurate measurement of the elements of $\chi^{\text{shg}}$ must be controlled for incident integrated pulse intensity (equivalently, energy), diameter, wavelength and duration as well as the extrinsic factor of the lens focal length. Since the value for $\chi_{ijk}^{\text{shg}}$ at \SI{800}{\nano\meter} is known from previous study \cite{wu2017giant}, all measurements were normalized to our measured response at \SI{800}{\nano\meter}. In addition, the relative enhancement of second harmonic response between TaAs and GaAs was checked independently using a two color Er:doped fiber laser with source photon wavelengths of 1550nm and 780nm respectively.

\subsection{Experimental measurement of correction factors}
\subsubsection{Intrinsic factors}

\begin{figure}\centering
\includegraphics[scale=0.8]{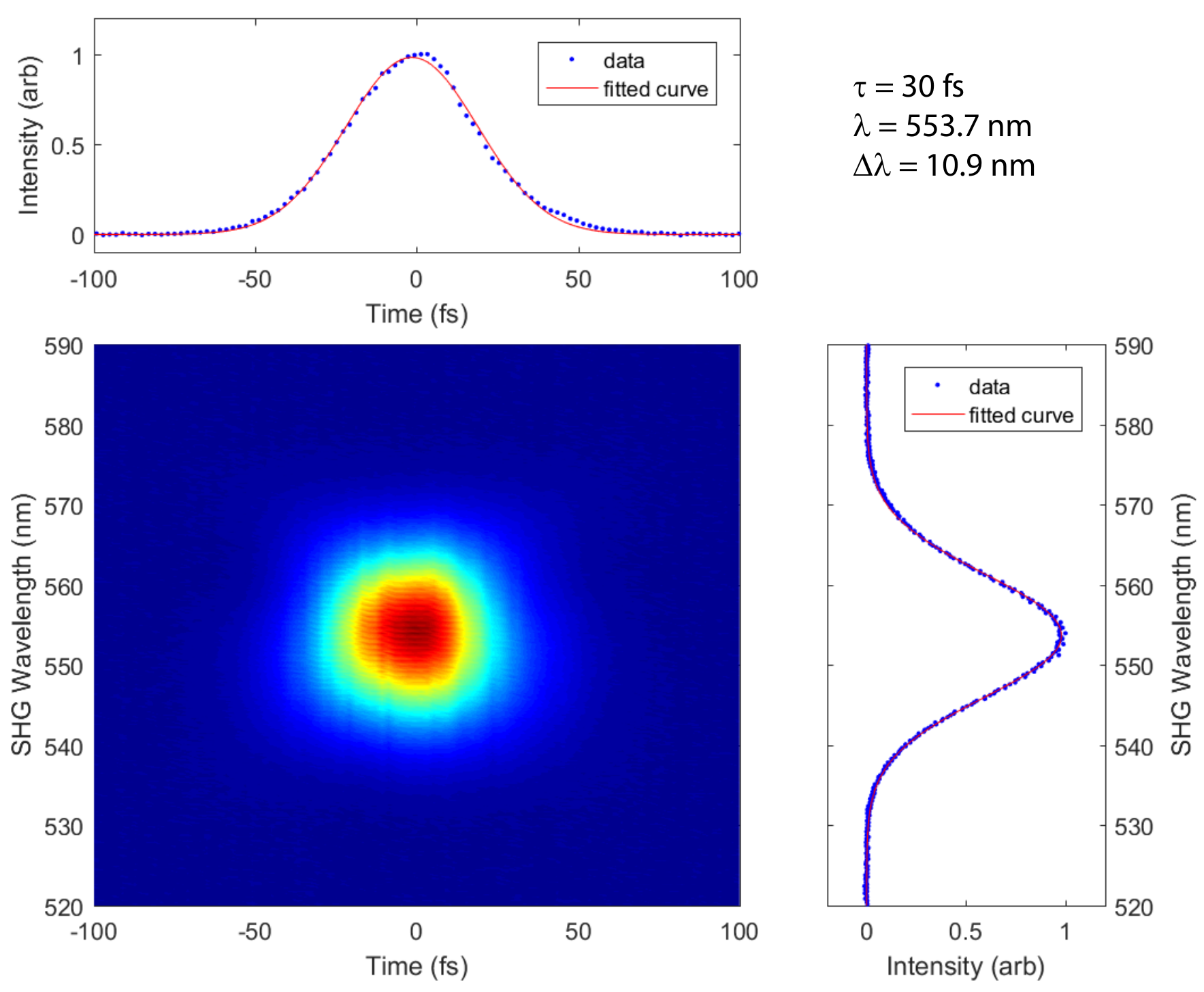}
\caption{Measured FROG spectrum for input fields $\lambda = 1100$~nm (nominally) showing the projection of the data along the temporal (above) and spectral (right) axes. Fits of the projected data to Guassian models, as described in the text, produced the observed pulse duration $\tau$, the center SHG wavelength $\lambda$ and the spectral width of the pulse $\Delta\lambda$.}\label{fig:sup_FROG}
\end{figure}

Even though a commercially supplied wavelength separator was employed, a number of parasitic wavelengths were present due to a combination of leakage of other responses from the Optical Parametric Amplifier, as well as their sum-frequency, difference-frequency, and second harmonic responses among the signal, idler and second stage pump fields. At each individual wavelength, we computed these parasitic wavelengths and assembled a combination of longpass and shortpass filters of minimum ND8 to remove the parasitics before measuring the incident power. This power was recorded after the last longpass filter (LF in figure 1 of the main text), and a set of reflective neutral density filters of value ND chosen to maintain a calculated incident fluence of roughly 
$\sim$ 20~mJ/cm$^2$.
As the measured average power $P$ is simply proportional to the energy per pulse through a factor of the repetition rate, the value $P/10^{ND}$ was divided from the measured power to provide a measure of the incident integrated intensity $I_\omega$. 

The spectral and temporal characteristics of the OPA beam were measured using a homebuilt Frequency Resolved Optical Gating (FROG) apparatus~\cite{} over the range from incident wavelength $\lambda = 800$~nm to $\lambda = 2000$~nm. In all cases, the FROG measurement was performed on the beam before its entrance into the reflective objective (i.e., in between LF and RO in figure 1 of the main text) in order to account for the filtering and dispersive characteristics of the preceding optics. While we did not use the same ND filters for the FROG measurement as we did in the experiments, we did use the same number of filters of identical material (UV grade fused silica) and thickness as those used in data collection. A representative FROG trace is shown in figure~\ref{fig:sup_FROG}. The FROG apparatus used a \SI{100}{\micro\meter} thick $\beta$-Barium Oxide (BBO) 
crystal, the radiated second harmonic spectrum was determined via simulation with lab2.de software\cite{Lab2} to be virtually identical to that which would derive from a thin layer of a bulk response reflection experiment as occured here on \ce{TaAs}. 

The pulse duration of the incident second harmonic field was determined from numerically summing the FROG trace along the spectral dimension and fitting the resulting data to a Gaussian of the form $I_{2\omega}(t) = I_0\exp\left[-\left((t-t_0)/\tau\right)^2\right]$, as shown in the top panel of figure~\ref{fig:sup_FROG}. We note that from evaluation of the convolution of a Gaussian with itself, the pulse duration at $2\omega$ is related to that at $\omega$ through a numerical factor as $\tau_{2\omega} = \tau_{\omega}/\sqrt{2}$. In order to determine the radiated central $2\omega$ wavelength and bandwidth, we numerically summed the FROG trace along the temporal dimension and fit the result to a Gaussian of the form 
$I_{2\omega}(\lambda) = I_0\exp\left[-\left((\lambda-\lambda_0)/\Delta\lambda\right)^2\right]$, as shown on the righthand panel. These data allowed us to determine the fundamental wavelength as $2\lambda_0$ and the spectral bandwidth of the SHG pulse $\Delta\lambda$.

\begin{figure}\centering
\includegraphics[width=0.99\textwidth]{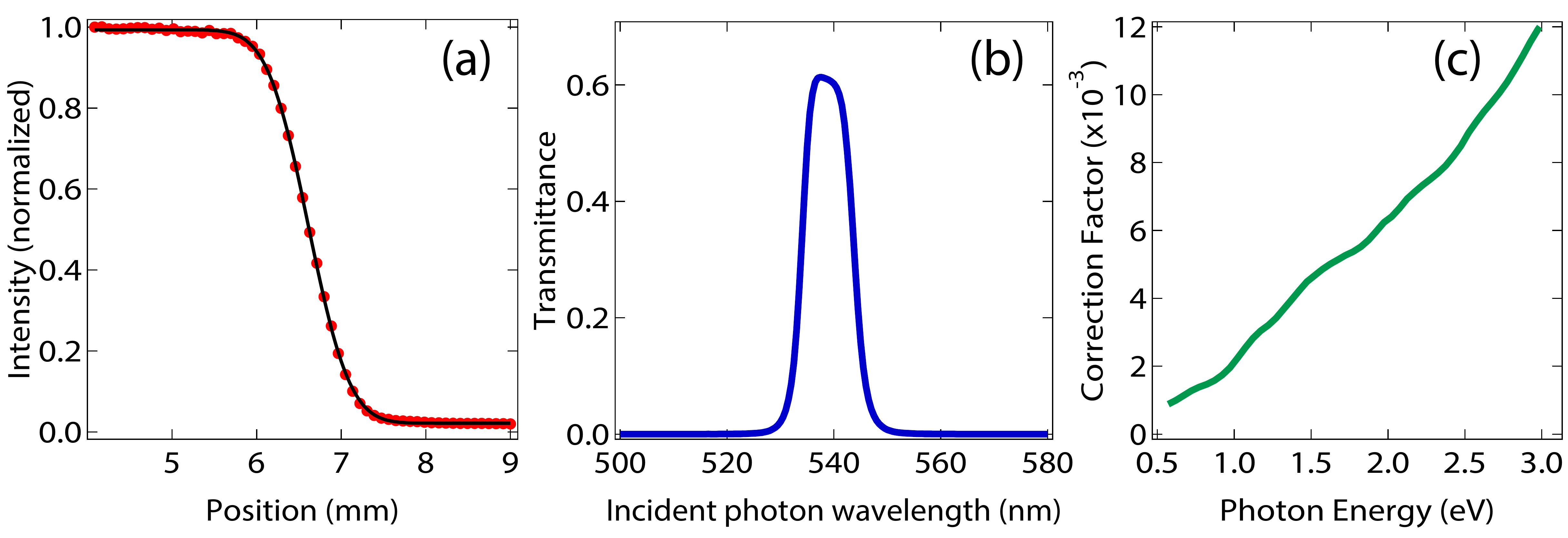}
\caption{(a) Representative data from knife-edge measurement of the beam diameter and corresponding fit for $\lambda_0=1100$~nm incident light. (b) Experimentally measured transmittance of the 540~nm bandpass filter. (c) The linear optical correction factor $BP(\omega)$ as a function of incident photon energy.
}
\label{fig:spot_transmittance_correction}
\end{figure}


The knife-edge method was used to measure the beam diameter $w_0$ 
immediately before the reflective objective as a function of incident wavelength $\lambda$ for all wavelengths in the study. Representative data at $\lambda = 1100$~nm and the corresponding fit to the complementary error function $I(x) = I_0\erfc{\left(\sqrt{2}(x-x_0)/w_0\right)}$ are shown in figure~\ref{fig:spot_transmittance_correction}. A basic Gaussian beam propagation could be used to determine the beam at the focus. However, this was unnecessary since the identical, all-reflective optical setup was used for every data point, necessitating only measurement of the beam diameter $w_0$.

\subsubsection{Extrinsic factors}

Experimentally, the measured second harmonic intensity is also proportional to the spectral response of the detector $\mathcal{D}(\lambda)$ and the transmittance $\mathcal{T}(\lambda)$ of the filters in front of the detector. 

The detector spectral response $\mathcal{D}(\lambda)$ was determined from the manufacturer specifications for each detector and divided out from the measured intensity. In the range from $\lambda = 800$~nm to $\lambda = 1600$~nm, we used a Hamamatsu R12829 in a Hamamatsu C12597-01 socket for power supply. The photomultiplier output was directly connected to the current input of a Zurich Instruments MFLI digitizing lockin which yielded a current value proportional to the incident SHG power. The range from $\lambda = 1560$~nm to $\lambda = 1700$~nm was measured with an unbiased Thorlabs FDS010 photodiode, while measurements from $\lambda = 1600$~nm to $\lambda = 2600$~nm performed with a Thorlabs FGA01 \ce{InGaAs} photodiode. For their respective wavelength ranges, the photodiodes were connected to the input of a Cremat CR-Z-110 charge preamplifier followed by a CR-S-8$\mu$s shaping amplifier whose output were connected to the voltage input of the lockin. The overlap between the wavelength ranges of the three detectors allowed us to account for the current to voltage conversion of the charge integrating electronics in relation to the PMT, permitting the \ce{Si} photodiode to be a bridge between the PMT and the \ce{InGaAs} detector.

In all cases, we used a set of two shortpass filters and a single bandpass filter centered at the nominal second harmonic wavelength, which determined the spectral transmittance $\mathcal{T}(\lambda)$. These filters were chosen in order to provide attenuation of $>$ND10 at the fundamental wavelength while not attenuating the SHG by more than a factor of 2. We verified the transmittance of the filters to match the supplier's specifications through individual UV-Vis measurements, as shown in figure~\ref{fig:spot_transmittance_correction}. It was necessary to account for the fact that the bandwidth of the filter was more narrow than that of the second harmonic pulse, in particular at the longest wavelengths measured. Thus, after dividing by $\mathcal{D}(\lambda)$, the true incident SHG power was determined by dividing out the numerically integrated product 
 of the measured filter data and a normalized Gaussian representing the fitted parameters of the measured second harmonic spectrum the FROG traces as

\begin{equation}
\frac{1}{\sqrt{\pi}\Delta\lambda}\int_{-\infty}^{\infty} \mathcal{T}(\lambda)\exp\left[-((\lambda-\lambda_0)/\Delta\lambda)^2\right]d\lambda.
\end{equation}

\subsection{Translating reflection geometry parameters to bulk nonlinear conductivity}
To calculate the bulk nonlinear optical parameters $\chi^{\text{shg}}$, and $\sigma^{\text{shg}}$, we also need to correct for the linear optical parameters of \ce{TaAs}. We use the formula derived by Bloembergen and Pershan\cite{bloembergen1962light,wu2017giant} to account for the linear optical properties such as the refractive index and transmission coefficients:
\begin{equation}
\chi_R^{\text{shg}}=\frac{\chi^{\text{shg}}}{\left(\sqrt{\epsilon(2\omega)} + \sqrt{\epsilon(\omega)}\right)\left(\sqrt{\epsilon(2\omega)}+1\right)} T(\omega)^2
\end{equation}
where $\chi_R^{\text{shg}}$ 
 denotes the nonlinear susceptibility in reflection geometry, as measured in the experiment, $\epsilon(\omega)$, and $T(\omega)$, are respectively the dielectric constant and the transmission coefficients of \ce{TaAs}.

We define the Bloembergen-Pershan correction coefficient $BP(\omega)\equiv\frac{T(\omega)^2}{\left(\sqrt{\epsilon(2\omega)} + \sqrt{\epsilon(\omega)}\right)\left(\sqrt{\epsilon(2\omega)}+1\right)}$, which can be calculated as a function of incident photon energy based on values of the dielectric constant as measured by ellipsometry.\footnote{The dielectric constant was measured using ellipsometry of \ce{TaAs} by Bing Xu at Universit\'e de Fribourg, and obtained in private communication.} (See figure \ref{fig:spot_transmittance_correction}.) Optical conductivity is more conveniently modeled than susceptibility, and can be obtained using the relation: $\sigma^{\text{shg}}_{ijk}(\omega)=-i\epsilon_0\,\omega\,\chi^{\text{shg}}_{ijk}(\omega)$



\section{Quasi-One Dimensional Model for Second Harmonic Generation}

\subsection{Model details}

%
%

This supplement describes the details of the phenomenological model of Rice-Mele chains used in the main text, and details the calculation of its second harmonic response. The model is inspired by the polar character of the \ce{TaAs} lattice as well as the observation that  $\n{\sigma_{zzz}}$ is the dominant component of the SHG response. This is reminiscent of a 1D system, in which $\sigma_{zzz}$ is the only non-zero component. 
We consider a quasi-one dimensional model composed of 1D Rice Mele chains in the $z$-direction~\cite{RM82} with only weak couplings in the $xy$ plane. 
In momentum space it takes the form of a two band model $H_{\v{k}} = \v{d}_{\v{k}} \cdot \v{\sigma}$ with
\begin{equation}
\begin{aligned}
d_x &= t  \sin(c k_z/2)\\
d_y &= \delta \cos(c k_z/2)\\
d_z &= \Delta +t_{AB}(\cos(b k_x)+\cos(b k_y)).
\end{aligned}
\label{eq:RM}
\end{equation}
Here $b$ and $c$ are the dimension of the unit cell in the $x$ and $z$ directions. Along lines in the $z$-direction there is a staggered onsite potential $\pm\Delta$ and staggered hopping $t  \pm \delta$. The parameter $t_{AB} = t_{\parallel,A} - t_{\parallel,B} \ll \Delta$ represents the difference between the inter-chain hopping strengths. When $t_{AB} = 0$, this reduces to an ensemble of independent Rice-Mele models. The overall energy scale is set by fixing $\Delta$; there are three independent parameters of the model: $\tilde{t}  = t /\Delta$, $\tilde{\delta} = \delta/\Delta$, and $\tilde{t}_{AB} = t_{AB}/\Delta$.

\subsection{Calculation of SHG Response}

The second-order conductivity tensor is defined via
\begin{equation}
	J_a(\omega_0) = \sum_{b,c} \sigma_{abc} (\omega_0; \omega_1, \omega_1) E_b(\omega_1) E_c(\omega_2)
\end{equation}
where $\omega_0 = \omega_1 + \omega_2$ is the frequency of the emitted photon The second harmonic response is $\sigma_{abc}^{\text{shg}}(\omega) = \sigma_{abc}(2\omega;\omega,\omega)$. We will also frequently invoke the shift current, $\sigma_{abc}^{\text{shift}}(\omega) = \sigma_{abc}(0;-\omega,\omega)$. The shift current may be thought of as the ``solar panel response'': the amount of DC current generated under illumination.

In two band models, the diagonal components of $\sigma_{abc}$ are particularly simple. For $\omega > 0$,\cite{MN16,yang2017divergent} 
\begin{align}
	\re \sigma_{aaa}^\text{shift}(\omega) \ &=\ \frac{2\pi e^3}{\hbar^2 \omega^2} \ibz f_{10} v_{01}^a w_{10}^{aa} \delta(\omega_{10} - \omega)\\
	\label{eq:shift_two_bands}
	\re \sigma_{aaa}^\text{shg}(\omega) \ &=\ \frac{\pi e^3}{2 \hbar^2 \omega^2} \ibz \Big[
	f_{10} v_{01}^a w_{10}^{aa} \delta(\omega_{10} - 2\omega)
	- 2f_{10} v_{01}^a w_{10}^{aa} \delta(\omega_{10} - \omega)
\Big]
\end{align}
where $\ibz = \int \frac{d^d\v{k}}{(2\pi)^d}$ is the normalized integral over the Brillouin zone, $0$ and $1$ refer to the valence and conduction bands respectively, $f_{01}$ is the difference in Fermi factors, $\omega_{01}$ is the difference in band frequencies, $v^a = \partial_{k_a} H$ is the velocity operator, and $w^{aa} = \partial_{k_a} \partial_{k_a} H$. Taken together, these lead to the convenient identity (see equation (3) of the main text)
\begin{equation}
	\re \sigma_{aaa}^\text{shg}(\omega) = - \frac{1}{2}\re \sigma_{aaa}^{\text{shift}}(\omega)  + \re \sigma_{aaa}^{\text{shift}}(2\omega).
	\label{eq:shift_SHG_relation_two_bands}
\end{equation}
Again, these equations only hold for two band models.

We now evaluate these equations for our specific model. We focus on the two-photon term of the SHG. Differentiating equation \eqref{eq:RM},
\begin{equation}
	\hat{v}^z = \frac{c}{\hbar 2} \left[ -t  \sin\left( \frac{k_z c}{2} \right) \sigma^x + \delta \cos\left( \frac{k_z c}{2} \right) \sigma^y \right]
\end{equation}
and
\begin{equation}
	\hat{w}^{zz} = \frac{c^2}{4\hbar} \left[ -t \cos\left( \frac{k_z c}{2} \right) \sigma^x - \delta \sin\left( \frac{k_z c}{2} \right) \sigma^y \right].
\end{equation}
Hence, using $\sigma^i \sigma^j = i \varepsilon^{ij}_k \sigma^k$,
\begin{equation}
	\hat{v}^z \hat{w}^{zz} = \frac{1}{\hbar^2} \left( \frac{c}{2} \right)^3 \left[ \frac{t^2 - \delta^2}{2} \sin\left( \frac{2 k_z c}{2} \right) + i t\delta \right] \sigma^z.
	\label{eq:matrix_element_I}
\end{equation}

We must now evaluate the matrix element $\braket{0|\sigma^z|0}$. Adopting the Bloch sphere representation,
\begin{equation}
	\v{d} = d \left( \sin \theta \cos \varphi, \sin \theta \sin \varphi, \cos \theta \right)
\end{equation}
which leads to the eigenvector equation $H \ket{0} = \varepsilon_0 \ket{0}$ with
\begin{equation}
	\ket{0} = \begin{pmatrix}
		\cos \frac{\theta}{2}\\
		\sin \frac{\theta}{2} e^{i\varphi}
	\end{pmatrix},\qquad
\end{equation}
and eigenvalue $\varepsilon_v = -d$. Therefore
\begin{equation}
	\braket{0|\sigma^z|0} = \left( \cos^2 \frac{\theta}{2} - \sin^2 \frac{\theta}{2} \right) = \cos \theta = \frac{d_z}{d}.
	\label{eq:matrix_element_II}
\end{equation}

Combining \eqref{eq:matrix_element_I} and \eqref{eq:matrix_element_II} shows that the two-photon contribution is
\begin{equation}
	\re \sigma_{zzz}^{\text{shg2p}}(\omega) =
	 \frac{\pi e^e}{2 \hbar^2 \omega^2} \ibz 	f_{10} v_{01}^z w_{10}^{zz} \delta(\omega_{01} - 2\omega)
=
	i \frac{\pi e^3}{2 \hbar^2 \omega^2}
	\ibz (-1) \frac{1}{\hbar^2}\left( \frac{c}{2} \right)^3 i t \delta \frac{d_z}{d} \delta(\omega_{10} - 2 \omega).
\end{equation}
Note that the $\sin(k_z c)$ term integrates to zero and can be ignored. What remains is to de-dimensionalize the integral and simplify the expression somewhat. Define $\varepsilon(k) = d(k) = \hbar \omega_{10}/2$. 
Then
\begin{equation}
	\re \sigma_{zzz}^{\text{shg2p}}(\omega)
		= \frac{\pi e^3}{2\hbar^2} \left( \frac{c}{2} \right)^3 t \delta \int \frac{d^3 \v{k}}{\left( 2\pi \right)^3} \frac{d_z}{ \varepsilon(k) \omega^2} \frac{\hbar}{2}\delta(\varepsilon(k) - \hbar \omega).
\end{equation}

To account for physical broadening of the peak, we use a Lorentzian $\delta(\varepsilon - \omega) \to \frac{1}{\pi} \frac{\gamma}{\left( \varepsilon - \omega \right)^2 + \gamma^2}$. Furtheremore, we convert the remaining factors of $\omega$ into $\varepsilon$ to remove an unphysical diverges at $\omega \to 0$ in numerical evaluation. So
\begin{equation}
	\re \sigma_{zzz}^{\text{shg2p}}(\omega)	= \frac{\pi e^3}{\hbar} \left( \frac{c}{2} \right)^3 t \delta \frac{1}{4\pi}\ibz \frac{d_z}{ \varepsilon(k)^3}  \frac{\gamma}{\left( \varepsilon(k) - \omega \right)^2 + \gamma^2}.
\end{equation}
Now we de-dimensionalize by normalizing energies to $\Delta$, and henceforth represent $\Delta$-normalizes quantities with a tilde. For example, $\tilde{t} \equiv t/\Delta$. Furthermore, let us define $X \equiv k_x b, Y \equiv k_y b, Z \equiv k_z c$. This gives 
\begin{align}
	\widetilde{\varepsilon} = \sqrt{\tilde{\delta}^2\cos(Z/2)^2  + \tilde{t}^2  \sin(Z/2)^2+ \widetilde{d_z}^2}
	\text{ and }
	\widetilde{d_z} = 1 + \tilde{t}_{AB} \left[ \cos(X) + \cos(Y) \right].
\end{align}
Thus 
\begin{align}	\sigma_{zzz}^{\text{shg2p}}(\omega)
	&= \frac{e^3}{\hbar} \left( \frac{c}{2} \right)^3 \frac{\widetilde{t} \widetilde{\delta}}{\Delta} \frac{1}{4} \frac{1}{(2\pi)^3 abc}\; 
	\int_{-\pi}^\pi dX \, dY \, dZ \, \frac{\widetilde{d}_z}{ \tilde{\varepsilon}(k)^3} \frac{\tilde{\gamma}}{\left( \tilde{\omega} - \tilde{\varepsilon}(k) \right)^2 + \tilde{\gamma}^2}.
\end{align}

To match with experimental data, one should multiply by degeneracy factors that account for the spin degeneracy $g_s = 2 $ and orbital degeneracy $g_O = 4$ for the number of Rice-Mele chains per unit cell. 
Rearranging, we arrive at the two-photon contribution to the SHG 
\begin{equation}
	\sigma_{zzz}^\text{shg2p}(\omega)  
= \left[\frac{e^2}{\hbar} \frac{e}{\Delta} \right]
\frac{c^2}{b^2} \left( \frac{1}{4\pi} \right)^3  F(\tilde{\omega}; \tilde{\delta}, \tilde{t} , \tilde{t}_{AB}, \tilde{\gamma})
\end{equation}
where the term in brackets has units of conductance per Volt, and the integral has been re-packaged as

\begin{align}
F(\tilde{\omega}; \tilde{\delta}, \tilde{t} , \tilde{t}_{AB}, \tilde{\gamma})
\nonumber
& = g_s g_O \frac{\tilde{\delta} \tilde{t}}{4}\int_{-\pi}^\pi dX \,
 \int_{-\pi}^\pi dY \,
 \int_{-\pi}^\pi dZ \;
 \frac{1 + \tilde{t}_{AB} \left[ \cos(X) + \cos(Y) \right]}{\left(\tilde{\delta}^2\cos(Z/2)^2  + \tilde{t}^2  \sin(Z/2)^2 + \left( 1 + \tilde{t}_{AB} \left[ \cos(X) + \cos(Y) \right] \right)^2 \right)^{3/2}}\\
 \ &\hspace{2em} \times\ 
 \frac{\tilde{\gamma}}{\left( \tilde{\omega} -  \left[\tilde{\delta}^2 \cos(Z/2)^2 + \tilde{t}^2  \sin(Z/2)^2 + \left( 1 + \tilde{t}_{AB} \left[ \cos(X) + \cos(Y) \right]  \right)^2  \right]^{1/2}\right)^2 + \tilde{\gamma}^2}.
\end{align}
By Equation \eqref{eq:shift_SHG_relation_two_bands}, $\re \sigma^\text{shift}_{zzz}(\omega) = -2 \sigma_{zzz}^\text{shg2p}(\omega)$ (see equation (2) of the main text).

It is worth examining the special case where $t_{AB} = 0$, i.e. an ensemble of coupled Rice-Mele chains. Here the integral becomes analytically tractable. Converting the Lorentzian back to a $\delta$ function,
\begin{align}
	&\int_0^\infty d\omega \; F\left(\widetilde{\omega}; \tilde{\delta}, \tilde{t} , \tilde{t}_{AB} \to 0, \tilde{\gamma} \to 0\right)\\
	\ &=\ \frac{\pi \Delta g_s g_O}{\hbar} \left( 2\pi \right)^2 \frac{\tilde{\delta} \tilde{t}}{4} \int_{-\pi}^\pi \frac{dZ}{\left( \tilde{\delta}^2 \cos(Z/2)^2 + \tilde{t}^2  \sin(Z/2)^2 + 1 \right)^{3/2}}\\
	\ &=\ g_s g_0 \frac{\Delta}{\hbar} \frac{2 (2\pi)^3 \tilde{\delta} \tilde{t}}{4}\frac{E_2\left( \frac{\tilde{\delta}^2 -\tilde{t}^2 }{1 + \tilde{\delta}^2} \right)}{\left( 1+ \tilde{t}^2  \right) \sqrt{1 + \tilde{\delta}^2}},
\end{align}
where $E_2$ is the complete elliptic integral of the second kind. Therefore, using the fact that the spectral weights of the shift current and two-photon contribution to the SHG are equal,
\begin{equation}
	\Sigma^\text{shift}_z = 
	2\Sigma^\text{shg 2p}_z =
	\left[ \frac{e^3}{\hbar^2} \right] \frac{c^2}{b^2} G(\tilde{\delta}, \tilde{t} ) \text{ where } G(\tilde{\delta}, \tilde{t} ) \equiv g_s g_O  \frac{\tilde{\delta} \tilde{t}}{8} \frac{E_2\left( \frac{\tilde{\delta}^2 -\tilde{t}^2 }{1 + \tilde{\delta}^2} \right)}{\left( 1+ \tilde{t}^2  \right) \sqrt{1 + \tilde{\delta}^2}}.
	\label{eq:rice_mele_maximum}
\end{equation}
The function $G$ is bounded and attains its global maximum at $G(\sqrt{2},\sqrt{2}) = \pi 3^{-3/2} \approx 0.604$ (see equation (4) of the main text).
This shows that the total spectral weight in the two-photon contribution is bounded, i.e. there is a maximum amount of SHG for Rice-Mele models.


\section{Gauge-Invariant Cumulants and a new sum rule}

This appendix reviews the theoretical machinery of gauge-invariant cumulants (GICs) that underlies the relationship between ground-state polarization distributions and the frequency-integrated nonlinear response described in the last part of the main text. We restrict the analysis to two-band tight-binding models, relevant for this work, to show that this connection can be made exact.
It generalizes the known relationships between the Berry connection and polarization, as well as the relation between the spread of Wannier functions with linear response. The last part of the supplement shows how this connection gives a guide to construct of models whose spectral weight exceeds the maximum spectral weight possible in the Rice-Mele model~\cite{RM82}.

\subsection{The polarization distribution}
Let us start with the macroscopic polarization of a solid \cite{souza2000polarization,zak1989berry, king1993theory, resta1994macroscopic, resta2007theory, PhysRevB.49.14202} and Kohn's theory of the insulating state \cite{kohn1964theory}. Consider a solid with $N$ electrons and $M$ nuclei in any dimension $d$. The macroscopic polarization operator is 
\begin{align}
	\widehat{\v{P}} = \frac{1}{V} \left( e\widehat{\v{X}} + q_\text{nuc} \widehat{\v{X}}_{\text{nuc}} \right),
	\label{eq:polarization_operator}
\end{align}
where $\widehat{\v{X}} = \sum_{i=1}^N \hat{\v{x}}_i$ is the center-of-mass position of the electrons (resp. $\widehat{X}_{\text{nuc}}$ of the nuclei). Within the clamped nuclei approximation, the nuclei do not move and we can set $\widehat{X}_{\text{nuc}} = 0$ by a choice of coordinates. The macroscopic polarization is the expectation $\braket{\widehat{ \v{P}}}$.

The GICs are a systematic way to extract further information from $\widehat{\v{P}}$ that we now exploit. Computing $\braket{\widehat{ \v{P}}} = \braket{\Psi|\widehat{\v{P}}|\Psi}$, averages over many different centers of charge. Following Souza \textit{et al}~\cite{souza2000polarization} we define the \textit{distribution} of those centers of charge, i.e. the spatial distribution of polarization, via
\begin{equation}
	p(\v{X}) = \Braket{\Psi|\delta( \widehat{\v{X}} - \v{X})|\Psi}.
\end{equation}
This should be interpreted as the probability that the center of charge is exactly at the position $\v{X}$. 
Different electronic properties of the solid have already been directly linked to $p(\v{X})$, including the Berry connection, localization length of Wannier functions, and the f-sum rule\cite{souza2000polarization}.We can now add the second-harmonic to that list. As a cautionary note, the polarization is distinct from other real-space distributions in solids; $p(\v{X})$ is an different object from the electron density, or the Wannier functions. The subtle relationship between these is discussed carefully by Souza \textit{et. al.} \cite{souza2000polarization}.

The great utility of this distribution is tempered by the indirect way it must be computed, a complication due to the fact that the position operator $\widehat{\v{X}}$ is ill-defined \cite{resta1998quantum}. This implies that, even knowing the exact eigenvectors of the system, $p(X)$ cannot be straightforwardly computed and, moreover, it is not immediately clear if $p(X)$ is even a well-defined distribution. A key result of Souza \textit{et al} is that $p(X)$ can be carefully defined, rendering its cumulants computable.

\subsection{Definition of gauge-invariant cumulants}

\begin{figure}
	\centering
	\includegraphics[width=\textwidth]{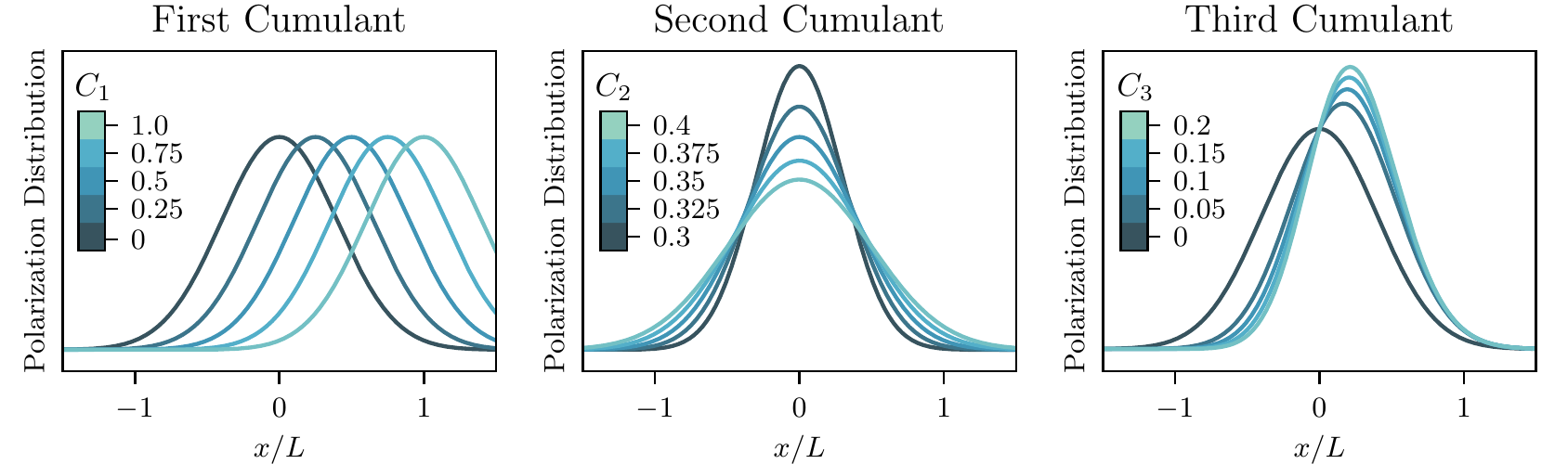}
	\caption{A schematic of the first three cumulants of the polarization distribution. In each panel, one cumulant is varied while the others are held fixed. Note that the true polarization distribution is periodic in $L$.}
	\label{fig:GIC_cartoon}
\end{figure}

Recall that the first several cumulants $C_n$ of a 
distribution $p(X)$ can be written in terms of the moments as
\begin{align}
	C_1^i  &= \braket{X^i}\\
	C_2^{ij}  &= \braket{X^i X^j} - \braket{X^i}\braket{X^j}\\
	C_3^{ijk}  &= \braket{X^iX^j X^k} + 2\braket{X^i}\braket{X^j}\braket{X^k}\\
	\notag
	&-\braket{X^i X^j}\braket{X^k}
	- \braket{X^i X^k}\braket{X^j}
	- \braket{X^j X^k}\braket{X^i}
	\label{eq:cumulants_from_moments}
\end{align}
where $i,j,k \in \st{x,y,z}$ are spatial indices. The first cumulant is the same as the first moment, the second is the variance, and the third cumulant is often called the skew  of the distribution. A schematic of the first three cumulants is shown in figure \eqref{fig:GIC_cartoon}.

A convenient way to compute the moments is in terms of the characteristic function
\begin{equation}
	\mathcal{C}(\v{\alpha}) = \braket{e^{-i\v{\alpha}\cdot \v{X}}},
\end{equation}
so that $\braket{X^i X^j \cdots X^k} = i \partial_{\alpha_i} i \partial_{\alpha_j} \cdots i \partial_{\alpha_k} C(\v{\alpha})\big|_{\v{\alpha}=0}$. 
The cumulants are likewise obtained by differentiating the log of the characteristic function 
\begin{equation}
	C_n^{ij\dots k} = i \partial_{\alpha_i} i \partial_{\alpha_j} \cdots i \partial_{\alpha_k} \ln \mathcal{C}(\v{\alpha})\big|_{\v{\alpha}=0}.
\end{equation}

Following \cite{souza2000polarization}, we define 
\begin{equation}
	\ln \mathcal{C}(\v{\alpha}) = \frac{V}{\left( 2\pi \right)^d} \ibz \; \ln \braket{\Psi_{\v{k}}|e^{-i\v{\alpha} \cdot \widehat{\v{X}}}|\Psi_{\v{k}+\v{\alpha}}}
	\label{eq:characteristic_function}
\end{equation}
where $V$ is the volume of the system, $\ibz$ denotes the normalized integral over the Brillouin zone, and $\ket{\Psi_k}$ is the many-body wavefunction with boundary conditions twisted by $e^{ikL}$.   
In a single-particle description, $\ket{\Psi_k} = \prod_{n} u_{\v{k} n} c^\dagger_{\v{k} n} \ket{0}$ where the Bloch wavefunctions $\psi_{\v{k}n}(\v{r}) = u_{\v{k}n}(\v{r}) e^{i\v{k}\cdot \v{r}}$ are chosen to satisfy a smooth, periodic gauge: $\psi_{\v{k}n} \equiv \psi_{\v{k} + \v{G},n}$ for reciprocal lattice vectors $\v{G}$. The cumulants of the polarization distribution can be computed via differentiating \eqref{eq:characteristic_function}. The first three are
\begin{align}
	C_1^i \ &=\ i V  \ibz \;\tr \left[c_1^i\right]\\
	C_2^{ij} \ &=\ i^2V  \ibz \;\tr \left[c^{ij}_2 - c_1^i c_1^j\right]\\
	C_3^{ijk} \ &=\ i^3V \ibz \; \tr \left[
		c_3^{ijk}  
		- c_2^{ij} c_1^{k} - c_2^{jk} c_1^i - c_2^{ki} c_1^j
+ 2 c_1^i c_1^j c_1^k
	\right]
\end{align}
where
the trace is over occupied bands, and the $c$'s are matrices in the band space whose matrix elements are
\begin{align}
	c_1^i &= i \braket{u_{kn}|\partial_{k_i} u_{km}},\\
	c_2^{ij} &= i^2 \braket{u_{kn}|\partial_{k_i} \partial_{k_j} u_{km}},\\
	c_3^{ijk} &= i^3 \braket{u_{kn}|\partial_{k_i} \partial_{k_j} \partial_{k_j} u_{km}}.
\end{align}
One can show that the cumulants are gauge-invariant--- hence the name --- and purely real in the presence of time-reversal symmetry.

To compute the cumulants numerically, it is advantageous to use an alternative formulation (written here for a single occupied band in 1D)\cite{hetenyi_cumulants}
\begin{equation}
	\Delta k  \prod_{i=0}^{M-1} \ln\braket{u_{k_i}|u_{k_{i+1}}} =  \sum_{n=1} \frac{(i \Delta k)^n}{n!} C_n
\end{equation}
where $k_{i+1} - k_i = \Delta k = 2\pi/M$, and again using a periodic gauge for the Bloch wavefunctions $\psi_{k}$. The cumulants can be measured by computing the left-hand side on successively finer meshes of $k$-points and fitting the result to a power series in $\Delta k$, the mesh size. The first several gauge-invariant cumulants in the Rice-Mele model have been computed recently, providing intuitive visualizations of the polarization distribution \cite{hetenyi_cumulants}.

\subsection{Interpretation and physical intuition}

Each of the GICs has a dual interpretation: as a measure of the spatial properties of the electrons, or as an electromagnetic response. The first cumulant tells us the mean of the center of charge polarization, i.e. the total deviation of electrons from their companion nuclei \cite{resta1994macroscopic}. Via equation \eqref{eq:polarization_operator}, the polarization of the solid has a straightforward formula in terms of the first cumulant 
\begin{equation}
	\braket{\widehat{\v{P}}} = \frac{e}{V} \v{C}_1 = e \sum_{n \text{ occ}} \ibz \v{A}_{\v{k}n},
	\label{eq:first_cumulant_polarization}
\end{equation}
where $\v{A}_{\v{k}n}$ is the Berry connection. The last expression is familiar from the modern theory of polarization. Equation \eqref{eq:first_cumulant_polarization} shows that, up to dimensionful prefactors, the first cumulant is the electrical response of the system in the absence of an external field, a purely static quantity.

The second cumulant measures the covariance of the polarization distribution. In an insulator, electrons are exponentially localized. Their localization is in the $i\; (=x,y,z)$ direction is given by (no summation)\cite{souza2000polarization}
\begin{equation}
	\xi_i = \lim_{N\to \infty} \sqrt{\frac{1}{N} C_2^{ii}}.
	\label{eq:localization_lengths}
\end{equation}
The second cumulant then gives a great deal of information about the character of the material. In a metal-insulator transition, for instance, the localization length diverges, and hence so does the second cumulant. Physically, the electrons in a metal are almost entirely delocalized, so the polarization distribution should be nearly flat, with variance on the order of the size of the system.

It is useful to contrast the information in $C_2$ with the Wannier functions. In 1d, the maximally localized Wannier functions have the property that the their centers are governed by the centers of polarization, and their variance is proportional to the squared localization length. We stress, however, that this does \textit{not} imply the Wannier density is the same as the polarization distribution. In dimensions greater than one, there are no unique maximally-localized single-particle Wannier functions, and all Wannier functions have variance strictly larger than the squared localization length. 

As an optical response, the second cumulant encodes the all-frequency linear response of the system. A standard application of the flucutation-dissipation theorem shows that the fluctuations in the polarization distribution ($C_2$) is related to the total current response of the system:\cite{souza2000polarization}
\begin{equation}
	\frac{ \pi e^2}{V^2 \hbar}	C_2^{\alpha\beta} = \frac{d\omega}{\omega} \re \sigma^{\alpha\beta}(\omega)
	\label{eq:f_sum_rule}
\end{equation}
where $\sigma^{\alpha\beta}$ is the linear conductivity. 

\subsection{A New Sum Rule}
Now that we have seen the dual nature --- spatial and optical --- of the first two GICs, it is perhaps not too surprising that the third cumulant can give a non-linear sum rule. The third cumulant measures the skewness of the polarization distribution which, in one dimension, says if the left or right ``shoulder'' of the distribution is larger.

On the optical side, we find that the second cumulant is related to the second-harmonic response of the system. Given that the first cumulant determines the polarization and the second cumulant gives a linear sum rule, it is perhaps not too surprising that the third cumulant can give a non-linear sum rule. More precisely, we show that for a generic two-band model (see equation (5) of the main text)
\begin{equation}
		\Sigma^\text{shift}_a = 	\frac{2\pi e^3}{V\hbar^2} C_3^a
\label{eq:third_cumulant_sum_rule}
\end{equation}
where
\begin{equation}
	\Sigma_a^\text{shift} = \int_0^\infty d\omega \; \re \sigma^{aaa}(0;-\omega,\omega)
\end{equation}
and, specializing to the case of two bands,
\begin{equation}
	C_3^a = -V\int \frac{d^d \v{k}}{(2\pi)^d} \im \left[ c_3 - 3c_1 c_2 + 2c_1^3 \right]
	\label{eq:C_3_two_band}
\end{equation}
where $c_n = \braket{0|\left( i \partial_{k_a} \right)^n|0}$ for the valence band Bloch wavefunction $\ket{0} = \ket{u_0(\v{k})}$.

Integrating \eqref{eq:shift_two_bands}, the spectral weight of the shift current in a two-band model is
\begin{align}
	\Sigma^{\text{shift}}_{a} &=  \frac{2\pi e^3}{\hbar^2}
\ibz \frac{\im[\bra 0 \partial_k H \ket 1 \bra 0 \partial_k^2 H \ket 0]}{(\epsilon_1-\epsilon_0)^2},
\label{eq:shift_spectral_weight}
\end{align} 
where $\ket 0$ and $\ket 1$ denote valence and conduction bands, respectively, and $\varepsilon_n = \varepsilon_n(k)$ are the band energies.

The Schr\"odinger equation and its $k$ derivatives give
\begin{align}
H \ket{n} &=  \epsilon_n \ket{n},\\
\partial_k H \ket{n} + H \ket{\partial_k n} &=  \partial_k \epsilon_n \ket{n} + \epsilon_n \ket{\partial_k n}, \\
\partial_k^2 H \ket{n} + \partial_k H \ket{\partial_k n} + H \ket{\partial_k^2 n} &=  \partial_k^2 \epsilon_n \ket{n} + \partial_k \epsilon_n \ket{\partial_k n} + \epsilon_n \ket{\partial_k^2 n}.
\end{align}

Taking inner products and applying $\ket{0}\bra{0} + \ket{1}\bra{1} = I$ implies
\begin{align}
\bra 0 \partial_k H \ket 1 &= (\epsilon_1 - \epsilon_0) \braket{ 0 | \partial_k 1}, \\
\nonumber
\bra 1 \partial_k^2 H \ket 0 &=(\epsilon_0 - \epsilon_1)[\braket{ 1 | \partial_k^2 0} - 2 \braket{ 1 | \partial_k 0} \braket{ 0 | \partial_k 0}] + 2 (\partial_k \epsilon_0 - \partial_k \epsilon_1) \braket{1 | \partial_k 0}.
\end{align}

Substituting these into the integral in the spectral weight yields
\begin{equation}
\begin{aligned}
&\frac{1}{(\epsilon_1-\epsilon_0)^2}
\im[\bra 0 \partial_k H \ket 1 \bra 0 \partial_k^2 H \ket 0]\\
&=\im\left\{\braket{ 0 | \partial_k 1}\left[- \braket{ 1 | \partial_k^2 0} + 2 \braket{ 1 | \partial_k 0} \braket{ 0 | \partial_k 0}] \}  - \frac{\partial_k \epsilon_0 - \partial_k \epsilon_1}{\epsilon_0 - \epsilon_1} 
\im[\braket{ 0 | \partial_k 1} \braket{1 | \partial_k 0}\right]\right\}.
\end{aligned}
\end{equation}
We can drop the last term since $\braket{ 0 | \partial_k 1} \braket{1 | \partial_k 0}$ is real.
Applying the resolution of the identity, the first term is
\begin{equation}
\begin{aligned}
\braket{ 0 | \partial_k 1}\braket{ 1 | \partial_k^2 0}
&= - \braket{ \partial_k 0 | \partial_k^2 0}
+ \braket{ \partial_k 0 | 0}\braket{ 0 | \partial_k^2 0} \\
&= - \partial_k(\braket{ 0 | \partial_k^2 0}) + \braket{ 0 | \partial_k^3 0} - \braket{ 0 | \partial_k 0}\braket{ 0 | \partial_k^2 0} \\
&= \partial_k c_2 + ic_3 -i c_1 c_2,
\end{aligned}
\end{equation}
while the second term becomes
\begin{equation}
\begin{aligned}
\braket{ 0 | \partial_k 1}\braket{ 1 | \partial_k 0} \braket{ 0 | \partial_k 0}
&= - \braket{ \partial_k  0 | \partial_k 0} \braket{ 0 | \partial_k 0}+ \braket{ \partial_k  0 | 0}\braket{ 0 | \partial_k 0} \braket{ 0 | \partial_k 0}\\
&= - \partial_k (\braket{ 0 | \partial_k 0})\braket{ 0 | \partial_k 0}+ \braket{ 0 | \partial_k^2 0} \braket{ 0 | \partial_k 0} - c_1^3\\
&= +\frac{1}{2}\partial_k(c_1^2) + i c_1 c_2 - i c_1^3.
\end{aligned}
\end{equation}
The total derivatives vanish after integration, and we obtain equation \eqref{eq:third_cumulant_sum_rule}.

This sum rule leads to intriguing conclusions. For two-band models, non-linear optical responses can be understood as a facet of the spatial distribution of polarization. This provides a clear physical picture that may be more intuitive than the normal expressions for SHG, which involve k-space sums over virtual transitions. 
Moreover, the sum rule can \textit{predict} the shift current --- and hence the SHG response --- of a material as a ground state property.

\subsection{Upper bounds in the Rice-Mele model and beyond}

In light of this relation between the SHG and $C_3$, it is worth revisiting the above fact, equation \eqref{eq:rice_mele_maximum}, that there is a maximum SHG in Rice-Mele models. While this maximum is a good benchmark as to how much second-harmonic can be produced by a system, it not an absolute bound. By designing a Hamiltonian with a large $C_3$, we will see this bound may be exceeded.

\begin{figure}
\centering
\includegraphics[width=\textwidth]{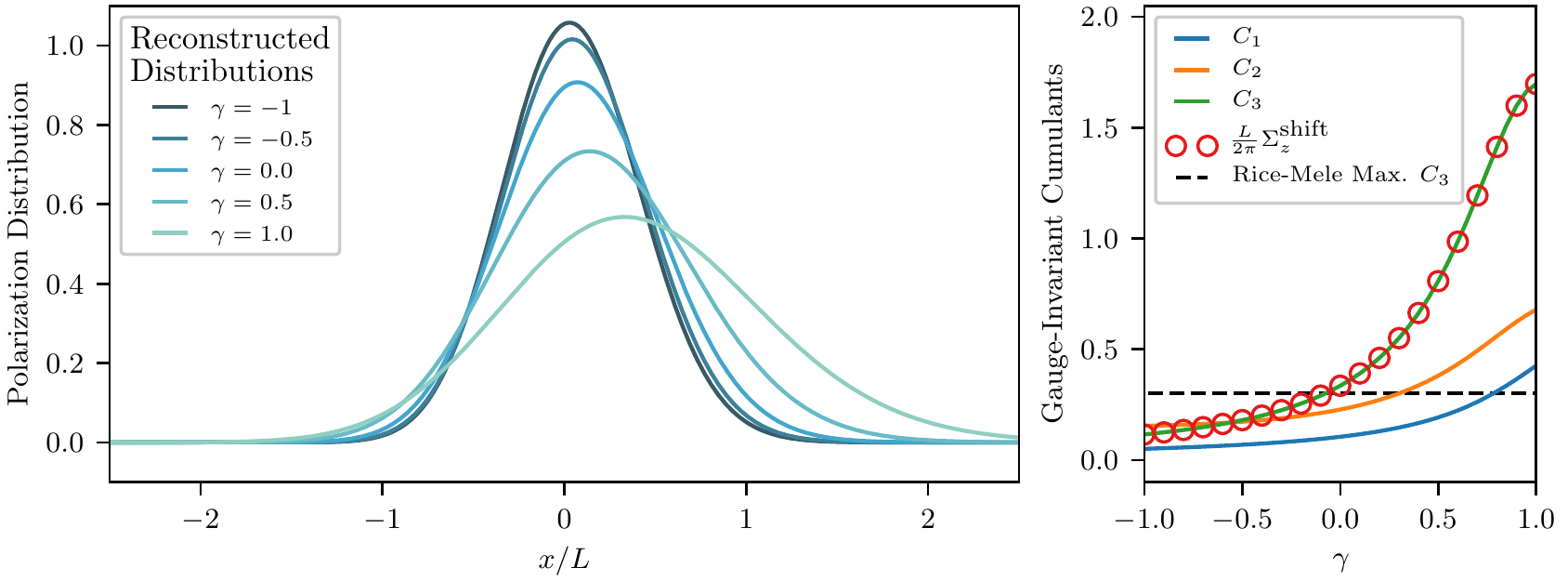}
\label{fig:bound_exceeded}
\caption{(Left) Reconstructed polarization distributions in the extended Rice-Mele model via the maximum entropy method\cite{hetenyi_cumulants}. (Right) The gauge-invariant cumulants and spectral weight in the next-nearest neighbor extension of the Rice-Mele model. The maximum in the normal Rice-Mele model, equation \eqref{eq:rice_mele_maximum}, is the black line, and the spectral weight as a function of $\gamma$ is computed via equation \ref{eq:shift_spectral_weight}. All parameters are given in the text.}
\end{figure}

For concreteness, consider a generalization of the Rice-Mele model with a next-nearest neighbor hopping $H = H_\text{RM} + H_\text{NNN}$ where
\begin{align}
	H_\text{RM} \ &=\ \sum_{n} \Delta (-1)^n c_{n}^\dagger c_n + \left( \frac{t}{2} + (-1)^n \frac{\delta}{2} \right) c^\dagger_n c_{n+1} + \text{h.c.}\\
	H_\text{NNN} \ &=\ \sum_n \left( \frac{t'}{2} + \frac{\delta'}{2} \right) c^\dagger_{An} c_{A,n+1}\\
	\ &\hspace{3em} + \left( \frac{t'}{2}  - \frac{\delta'}{2} \right) c^\dagger_{Bn} c_{B,n+1} + \text{h.c.}
\end{align}
In $k$-space this becomes
\begin{equation}
	H = \sum_k \v{c}_k^\dagger\left[ h_\text{RM}) + h_{\text{NNN}} \right] \v{c}_k 
	\label{eq:H_k_space}
\end{equation}
where
\begin{align}
	h_\text{RM}(\v{k})
	&=  
	t \cos\left( \frac{kc}{2}\right) \sigma_x +
	\delta \sin\left( \frac{kc}{2} \right) \sigma_y +
	\Delta \sigma_z\\ 
	h_\text{NNN}(\v{k}) 
	&= t'\cos(kc) \operatorname{Id}_2 + \delta' \cos(ka) \sigma_z.
\end{align}

By applying the maximum entropy method to approximately solve the moment problem \cite{hetenyi_cumulants}, we can reconstruct the polarization distribution. Figure~\ref{fig:bound_exceeded} shows the reconstructed polarization distributions in the next-nearest neighbor model with parameters $\Delta=1$, $\delta = \sqrt{2}$, $t = \sqrt{2} - 2\gamma$, $\delta'=\gamma$ and $t' = 0$. As $\gamma$ is tuned past zero, $C_3$ increases, and exceeds the maximum possible in the Rice-Mele model. We can thus tune the model to achieve an arbitrary large $C_3$ and, at a metal-insulator transition, cause it to diverge. Finding materials realizing a large third cumulant would be excellent candidates for giant shift current or second harmonic generation.

\bibliographystyle{naturemag}
\bibliography{nonlinear}

\end{document}